\documentclass[%
 reprint,
 amsmath,amssymb,
 aps,
pra,
floatfix,
]{revtex4-2}

\usepackage{graphicx}
\usepackage{dcolumn}
\usepackage{xcolor}

\usepackage{bm,physics}
\usepackage{lipsum}
\usepackage{xcolor}
\usepackage{multirow}
\usepackage{svg}
\usepackage{appendix}
\usepackage{mathtools}
\usepackage{makecell}
\usepackage{amsmath}

\usepackage{accents}
\begin{document}

\preprint{APS/123-QED}

\title{Non-Gaussian states via pump-depleted SPDC}

\author{Colin Vendromin}
\email{colin.vendromin@utoronto.ca}
\affiliation{Department of Physics, University of Toronto, 60 St. George Street, Toronto, Ontario, Canada M5S 1A7}%

\author{Samuel E. Fontaine}
\affiliation{Department of Physics, University of Toronto, 60 St. George Street, Toronto, Ontario, Canada M5S 1A7}%

\author{J. E. Sipe}
\affiliation{Department of Physics, University of Toronto, 60 St. George Street, Toronto, Ontario, Canada M5S 1A7}%

\date{\today}

\begin{abstract}
We develop a model for non-Gaussian state generation via spontaneous parametric down-conversion (SPDC) in InGaP microring resonators.
The nonlinear Hamiltonian is written in terms of the asymptotic fields for the system, which includes a phantom channel to handle scattering loss. The full ket for the system is written as a Gaussian unitary acting on a residual non-Gaussian ket, which is vacuum initially and evolves according to a non-Gaussian Hamiltonian. We show that for realistic parameters we can access the pump depletion regime, where the Wigner function for the residual non-Gaussian ket has negativity. 
But we find that the non-Gaussian features for the full ket could be 
unobservable due to the large amount of squeezing required to lead to pump depletion.
We show that a potential solution in the low-loss regime is to implement
an inverse Gaussian unitary on the accessible modes to remove most of the squeezing and reveal the non-Gaussian features.  This work provides a foundation for modeling pump-depleted SPDC in integrated lossy microring resonators, opening a path toward a scalable on-chip non-Gaussian source.
\end{abstract}

\maketitle


\section{Introduction}
Spontaneous parametric down-conversion (SPDC) is a widely used process in which a high-energy photon is converted into a pair of lower-energy photons via the second-order nonlinear susceptibility ($\chi^{(2)}$) of a medium. Heralded single photons \cite{HongMandel1986}, entangled photon pairs \cite{Kwiat1995}, and squeezed states of light \cite{Wu1986, schnabelSqueezeStates} can all be generated using SPDC, with applications across many quantum technologies, including quantum communication \cite{Gisin2007, Halder2008}, quantum key distribution \cite{Gisin2002, Zhong2015}, and continuous-variable quantum computing \cite{universal_qc_cv_cluster_menicucci_2006, YokoyamaClusterstate2013}.

Indium gallium phosphide (InGaP) is a promising material to enhance SPDC efficiency  due to its large $\chi^{(2)}$ coefficients \cite{Thiel2024}. For example, InGaP microring resonators have recently demonstrated high conversion efficiencies on the order of $10^5\,\%/{\rm W}$ with  on-chip power on the order of $1\,\mu{\rm W}$ \cite{Zhao2022, Akin2024}. Higher efficiency enables stronger squeezing and the potential for non-Gaussian states generated by pump depletion \cite{Florez2020, Jankowski2024, Yanagimoto2024}. Non-Gaussian states are a key resource for continuous-variable quantum computing, but they are typically generated probabilistically \cite{AghaeeRad2025}. Achieving a deterministic non-Gaussian source in a scalable, on-chip source would represent a crucial step toward scalable quantum computing architectures.

The combination of low pump power and high efficiency in InGaP microring resonators could allow 
access to the pump-depletion regime. Pump depletion from SPDC has already been observed in periodically poled lithium niobate (PPLN) waveguide systems \cite{Florez2020}. In these systems the generated photon number saturates in the depletion regime, and deviates from the exponential growth expected for purely Gaussian states. These observations suggest the generation of non-Gaussian states, consistent with theoretical modelling
\cite{Yanagimoto2022}. Demonstrations of pump depletion from second harmonic generation have also been demonstrated in a number of lithium niobate  platforms, including PPLN \cite{Park2022} and thin-film lithium niobate (TFLN) \cite{Zhao2020Lithium} waveguides, and in lithium niobate microrings \cite{du2024}. 

Recent theoretical work has begun to characterize the non-Gaussian states that emerge in the pump-depletion regime with models focused on PPLN waveguides \cite{Yanagimoto2022}. In these models about $-20\,{\rm dB}$ of squeezing is predicted from SPDC with only 200 pump photons, corresponding to roughly 10$\%$ pump depletion for a $1.4\,{\rm cm}$ crystal length, assuming advanced dispersion engineering. In such cases non-Gaussian features in the Wigner function appear along with the Gaussian ones characterizing  
squeezing, but if the squeezing is too large the non-Gaussian features are overwhelmed
\cite{Yanagimoto2024}, which is often the case when generating non-Gaussian states via pump depletion. Compared to microring models, waveguide models seem to achieve pump depletion with lower amounts of squeezing. However, even this lower amount of squeezing (around $-20$ dB) obscures the non-Gaussian features in the Wigner function, practically rendering these features inaccessible.
In addition, waveguide sources lack the compactness and scalability of integrated microrings. Theoretical work on the pump-depletion regime in microring resonators remains largely unexplored, particularly when accounting for multiple modes and scattering loss.

Reaching the depletion regime in microring models requires substantially more pump photons and higher squeezing than in the previous PPLN waveguide model \cite{Yanagimoto2022}. Higher squeezing makes the Gaussian features dominate, and the
non-Gaussianity becomes even harder to detect. Instead of relying on sources with lower squeezing to observe the non-Gaussianity, when scattering losses are sufficiently small a possible strategy would be to employ
a squeezed state interferometer \cite{Caves2020}. 
This would effectively apply an inverse Gaussian unitary on the accessible modes, removing the Gaussian features from the state and isolating its interesting non-Gaussian features. 

In this work we develop a multimode model of SPDC in InGaP microring resonators that is capable of describing non-Gaussian state generation in the pump-depletion regime. Our primary focus is on InGaP, but the model we present is readily adaptable for different $\chi^{(2)}$ materials. The Hamiltonian  is written in terms of the asymptotic fields for the system, which includes a phantom channel to describe
scattering loss \cite{Liscidni2012, Quesada2022}.  The full ket for the system is written as a Gaussian unitary acting on a residual non-Gaussian ket \cite{Yanagimoto2022}, where we derive an effective Hamiltonian that governs the evolution of the non-Gaussian ket. By writing the effective Hamiltonian in terms of supermodes, we are able to perform efficient numerical simulations for the non-Gaussian ket with only a few dominant supermodes. 

Using realistic parameters we calculate the squeezing level of the signal and idler fields, the amount of pump-depletion, and the Wigner function for the non-Gaussian ket. We show that under high levels of squeezing the non-Gaussian features in the Wigner function for the full ket indeed become obscured,
but that if the scattering loss is sufficiently small the non-Gaussian features can be restored by applying an inverse Gaussian unitary on the accessible modes. This unitary effectively \emph{stretches} the full ket to remove the squeezing, isolating the non-Gaussian features.

This paper is organized as follows. In Sec \ref{sec:Hamiltonian} we introduce the fields and Hamiltonian for the microring resonator system, including loss. In Sec. \ref{sec:SE solution} we then present
the solution of the Schr\"odinger equation for the full ket for the system, where we separate its evolution into Gaussian and non-Gaussian contributions. In Sec. \ref{sec:results} we first present results for the Gaussian part of the ket, including squeezing and the conversion efficiency of signal and idler photons, followed by results for the Wigner function of the non-Gaussian ket. We conclude that section with a discussion of how to isolate the non-Gaussian features of the full ket when there is loss. Finally, in Sec. \ref{sec:conclusion} we finish with 
a conclusion and outlook.

\section{Fields and Hamiltonian}
\label{sec:Hamiltonian}
 In this section we describe the fields of our system and the Hamiltonian that describes the SPDC interaction in the microring resonator. 

\begin{figure}[htb]
    \centering
    \includegraphics[width=0.8\linewidth]{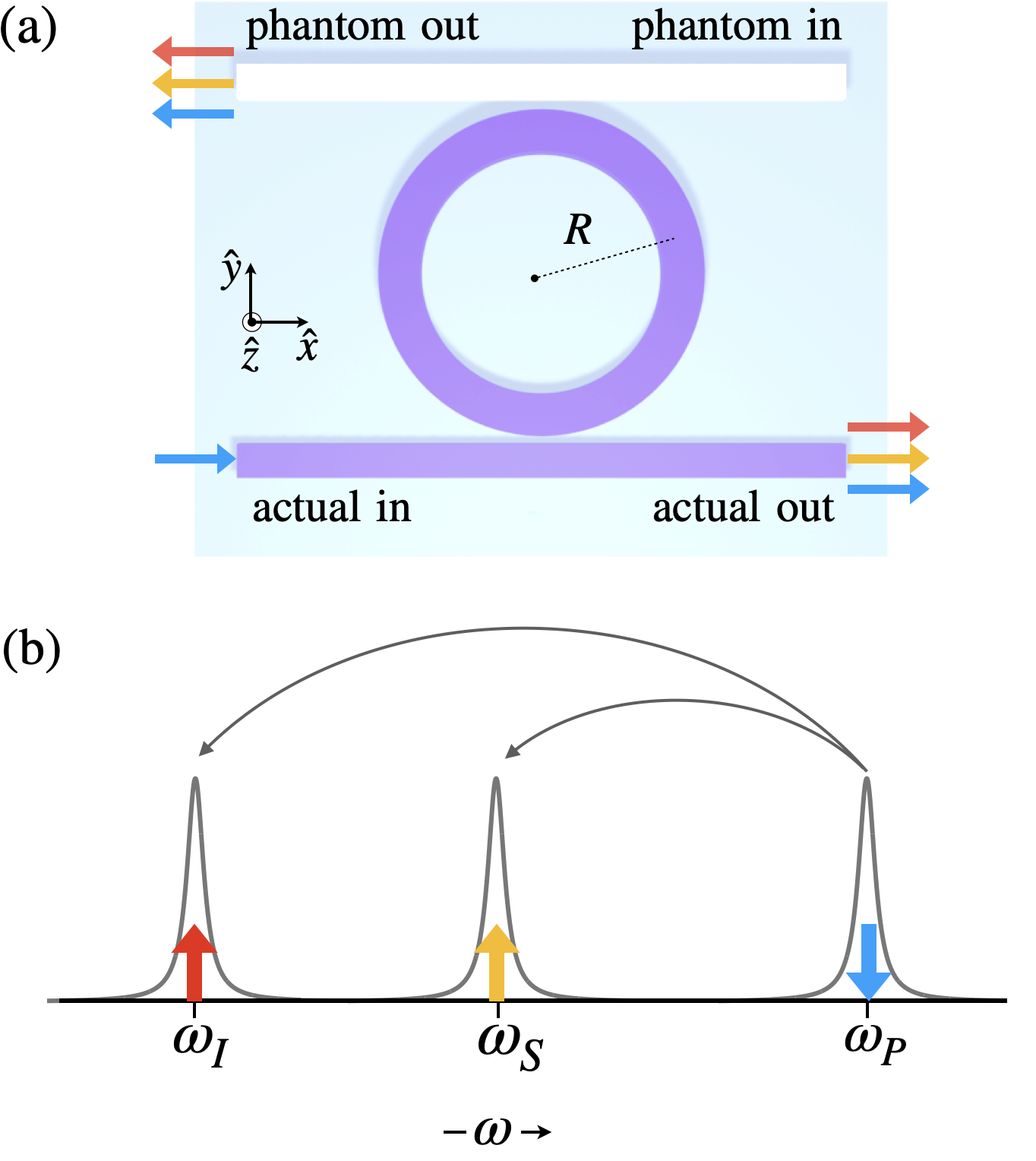}
    \caption{(a) Schematic of a ring resonator point-coupled to an actual waveguide (bottom purple) and phantom waveguide (white). (b) Schematic of the positions of the ring resonances for the SPDC interaction.}
    \label{fig:system}
\end{figure}

A schematic of the system we envision is shown in Fig. \ref{fig:system}(a). The actual waveguide and ring (purple) are made from ${\rm In}_{0.49}{\rm Ga}_{0.51}{\rm P}$ with thickness $102\,{\rm nm}$ and width $1194\,{\rm nm}$, where the dimensions are chosen such that a set of resonances are quasi-phase matched \cite{Fontaine2025}.  At an early time a pump pulse injected into the actual input channel propagates towards the ring. At $t=0$ the peak of the pump pulse reaches the coupling point, where it enters the ring and generates signal and idler photons via SPDC; we 
include loss by introducing
a phantom output channel that takes 
scattered photons out of the ring. The three ring resonances involved in the SPDC interaction are shown in Fig. \ref{fig:system}(b), where the center wavelengths of the pump ($P$), signal ($S$), and idler ($I$) resonances are $\lambda_P = 780\,{\rm nm}$, $\lambda_S = 1557.85\,{\rm nm}$, and $\lambda_I = 1562.45\,{\rm nm}$, respectively, where $\omega_J = 2\pi c/\lambda_J$ for $J \in \{P,S,I\}$.

We assume that the SPDC interaction is significant only in the ring resonator, where the fields are enhanced, and we formulate it in a manner
similar to previous work \cite{Liscidni2012} by borrowing a strategy from scattering theory. At $t=-\infty$ there is no nonlinear interaction and the only light in the system is a pump pulse moving towards the ring in the actual input channel. To describe the dynamics of the pump pulse we use a basis of asymptotic-in fields.  Similarly, at $t= \infty$ there is no nonlinear interaction; however, 
now the only light in the system is moving away from the ring in the actual and phantom output channels. To describe the dynamics of the generated signal and idler pulses we use a basis of asymptotic-out fields.

The displacement fields of the asymptotic-in and -out fields
are identified by a resonance label $J$, a channel label $n ={\rm ac}$ or $n={\rm ph}$ for the actual channel or phantom channel, respectively, and a continuous label $k$ for the wavenumber in either the actual or phantom waveguide. 
The positive frequency part of the displacement field operator for the pump $\bm D^+_P(\bm r)$ can be written as \cite{Liscidni2012}
\begin{align}
\label{eq:D+P}
\bm D^+_P(\bm r) &= \sum_{n={\rm ac, ph}}\int_0^\infty d k\,  b_{nPk} \bm D^{\rm in}_{nP k}(\bm r),
\end{align}
where $b_{nPk}$ destroys a pump photon in the asymptotic-in field for the input channel $n$ with wavenumber $k$, and $\bm D^{\rm in}_{nP k}(\bm r)$ is the asymptotic-in mode field associated with that photon. The pump operators satisfy the commutation relations
\begin{align}
\label{eq:b commute}
    [b_{nP k}, b^\dagger_{n P k'}] = \delta(k-k').
\end{align}
The positive frequency part of the displacement field operator for the signal and idler $\bm D^+_J(\bm r)$, for $J\in\{S,I\}$ can be written as 
\begin{align}
\label{eq:D+J}
\bm D^+_J(\bm r) &= \sum_{n={\rm ac, ph}}\int_0^\infty d k\,  a_{nJk} \bm D^{\rm out}_{n J k}(\bm r),
\end{align}
where $a_{nJk}$ destroys a signal ($J=S$) or idler ($J=I$) photon in the asymptotic-out field for the output channel $n$ with wavenumber $k$, and $\bm D^{\rm out}_{n J k}(\bm r)$ is the asymptotic-out mode field associated with that photon. The signal and idler operators satisfy the commutation relations
\begin{align}
\label{eq:a commute}
    [a_{n J k}, a^\dagger_{n J k'}] = \delta(k-k').
\end{align}
 
 The linear Hamiltonian can be written in terms of the asymptotic field operators as
\begin{align}
\label{eq:HL}
H_{\rm L} &=\sum_{n}\int_0^\infty d k \hbar \omega_{ n P k}b^\dagger_{ n P k} b_{nP k}\nonumber
\\
&+ \sum_{n, J\in\{S,I\}}\int_0^\infty d k \hbar \omega_{ nJ k}a^\dagger_{ n J k} a_{nJ k}, 
\end{align}
where $\omega_{nJk}$ is the dispersion relation for the actual ($n={\rm ac}$) or phantom ($n={\rm ph}$) channel for frequencies within the resonance labeled by $J$. We expand the dispersion relation in the channel $n$ around a reference wavenumber $K_{nJ}$ associated with the resonance $J$, neglecting
group velocity dispersion of frequencies within the resonance linewidths
\begin{align}
\label{eq:linear dispersion}
\omega_{nJk}\approx \omega_{J} + v_{nJ}(k - K_{nJ}),
\end{align}
where $\omega_J$ is the center frequency of the resonance $J$ and $v_{nJ}$ is the group velocity in the channel $n$ evaluated at the center frequency. The full dispersion relation, however, is used to calculate the center frequencies of the ring resonances. Over the frequencies spanning the 
multiple free-spectral-ranges
group velocity dispersion can be important,
and its effects are
captured in our model.  

The second-order nonlinear Hamiltonian for the SPDC interaction can be written as \cite{Quesada2022}
\begin{align}
\label{eq:HNL}
    H_{{\rm NL}} &= -\frac{2}{\epsilon_0}\int d\bm r \Gamma^{ijk}_2(\bm r) D^{-i}_S(\bm r) D^{-j}_I(\bm r)D^{+k}_P(\bm r) +{\rm H.c.},
\end{align}
where $\bm D^{-}_J(\bm r) = [\bm D^{+}_J(\bm r)]^\dagger$, $i$, $j$, and $k$ are Cartesian components and the tensor $\Gamma^{ijk}_2(\bm r)$ is related to the more familiar second-order tensor $\chi^{ijk}_2(\bm r)$ by \cite{Quesada2022}
\begin{align}
\label{eq:Gamma3}
    \Gamma^{ijk}_2(\bm r) &= \frac{\chi^{ijk}_2(\bm r)}{\epsilon_0\epsilon(\bm{r};\omega_P)\epsilon(\bm{r};\omega_S)\epsilon(\bm{r};\omega_I)},
\end{align}
where $\epsilon(\bm{r};\omega_J)$ is the dielectric constant at position $\bm{r}$ for frequency $\omega_J$ \cite{Quesada2022}.
Putting Eqs. \eqref{eq:D+P} and \eqref{eq:D+J} for the pump displacement field operator and the signal and idler displacement field operators respectively into Eq. \eqref{eq:HNL} for the nonlinear Hamiltonian, and only considering the case where a pump photon is annihilated from the actual input channel, we obtain
\begin{align}
    \label{eq:HSPDC continuous}
    H_{\rm NL} &= \sum_{n_1, n_2} \int dk_1dk_2dk_3 \hbar\Lambda^{{\rm ac}\, n_1 n_2}(k_1,k_2,k_3)  \nonumber
    \\
    &\times a^\dagger_{n_1 S k_1}a^\dagger_{n_2 I k_2}b_{{\rm ac}Pk_3} + {\rm H.c.},
\end{align}
where $\Lambda^{{\rm ac}\, n_1 n_2}(k_1,k_2,k_3)$ is the SPDC nonlinear coefficient obtained by integrating the field profiles for positions inside the ring resonator. It can be written as \cite{Fontaine2025}
\begin{align}
   \Lambda^{{\rm ac}\, n_1 n_2}(k_1,k_2,k_3) &= \overline{\chi}_2 R \sqrt{\frac{\hbar \omega_S\omega_I\omega_P}{4\pi \epsilon_0 A_{\rm eff}}}F^*_{n_1 S+}(k_1)\nonumber
    \\
    &\times F^*_{n_2 I+}(k_2)F_{{\rm ac} P-}(k_3),
\end{align}
where $\overline{\chi}_2 = 220\,{\rm pm /V}$ \cite{Ueno:97} is a typical value of the second-order tensor for InGaP, $R = 30\,\mu{\rm m}$ is the ring radius, and $A_{\rm eff} = 0.56\,(\mu{\rm m})^2$ is the effective area \cite{Fontaine2025}. We have introduced the Lorentzian field-enhancement factors \cite{MBanicRingLoss2022}
\begin{align}
    F_{nJ\pm}(k)&=\sqrt{\frac{v_{nJ}\overline{\Gamma}_J\eta_J}{\pi R}}\frac{1}{v_{nJ}(K_{nJ} - k) \pm i\overline{\Gamma}_J},
\end{align}
where $2\overline{\Gamma}_J$ is the linewidth of the resonance and $\eta_J$ is the escape efficiency into the actual channel, defined as
\begin{align}
    \eta_J \equiv 1- \frac{Q_{{\rm load},J}}{Q_{{\rm int}, J}},
\end{align}
where $Q_{{\rm load},J} = \omega_J/(2\overline{\Gamma}_J)$ is the loaded quality factor and $Q_{{\rm int}, J}$ is the intrinsic quality factor due to scattering loss. For InGaP we adopt realistic intrinsic quality factors of $Q_{{\rm int}, S} = Q_{{\rm int}, I} = 10^6$ and  $Q_{{\rm int}, P} = 10^5$ \cite{Akin2024, Thiel2024}.

We discretize the continuous wavenumbers to perform numerical calculations \cite{Quesada2022}, putting
for example $\int dk_1 \rightarrow \Delta k \sum_{k_1}$, where $\Delta k$ is the spacing between the discrete wavenumbers. Then
Eq. \eqref{eq:HSPDC continuous} for $H_{\rm NL}$  can be written as
\begin{align}
    \label{eq:HSPDC}
    H_{\rm NL} &= \sum_{\mu_1,\mu_2, \mu_3} \hbar[\Lambda]_{\mu_1 \mu_2 \mu_3}  a^\dagger_{S\mu_1}a^\dagger_{I\mu_2}b_{P\mu_3} + {\rm H.c.},
\end{align}
where we have introduced Greek indices that group together a channel and discrete wavenumber as
\begin{align}
\label{eq:greek indices}
    \mu_1 & = (n_1,k_1), \,\,\, \mu_2  = (n_2,k_2),\,\,\, \mu_3  = ({\rm ac},k_3),
\end{align}
where $k_i$ for $i\in\{1,2,3\}$ are the discrete wavenumbers, and we have defined
\begin{align}
    \label{eq:discrete nonlinear parameter}
    [\Lambda]_{\mu_1 \mu_2 \mu_3} \equiv (\Delta k)^{3/2}\Lambda^{{\rm ac}\, n_1 n_2}(k_1,k_2,k_3),
\end{align}
which has units of frequency.
The discrete operators in Eq. \eqref{eq:HSPDC} satisfy the commutation relations
\begin{align}
    [a_{J\mu}, a^\dagger_{J\mu'}]= [b_{P\mu}, b^\dagger_{P\mu'}] &= \delta_{\mu\mu'},
\end{align}
and all others are zero.

\section{Solution to the Schr\"odinger equation}
\label{sec:SE solution}
In this section we derive the dynamics of the ket
that evolves when
Eq. \eqref{eq:HSPDC} is used for the SPDC Hamiltonian.

We begin by considering the state evolution generally within an asymptotic fields analysis
\cite{Liscidni2012, Vendromin2024}. The state of the system $\ket{\Psi(t)}$ evolves according to the Schr\"odinger equation 
\begin{align}
\label{eq:SE full state}
    i\hbar \frac{d\ket{\Psi(t)}}{dt} = (H_{\rm L}+H_{\rm NL})\ket{\Psi(t)},
\end{align}
where $H_{\rm L}$ and $H_{\rm NL}$ are given by Eqs. \eqref{eq:HL} and \eqref{eq:HSPDC continuous} respectively. We assume that initially the pump pulse is in a coherent state in the actual channel evolving according to $H_{\rm L}$ far from the ring. At this early time $t = t_0<0$ we can put $H_{\rm NL} = 0$ in Eq. \eqref{eq:SE full state}, and write the initial state for the system as a coherent state for the pump, and vacuum for the signal and idler,
\begin{align}
\label{eq:initial state}
    \ket{\Psi(t_0)} = \exp(\int dk\, x(k,t_0) b^\dagger_{{\rm ac}Pk} - \rm H.c.)\ket{\rm vac},
\end{align}
where $x(k,t_0)$ is the initial coherent state amplitude in terms of the wavenumbers for the channel. Here $x(k,t_0)$ is given by
\begin{align}
\label{eq:initial pulse}
    x(k,t_0) = \sqrt{N_P}\Phi(k) {\rm e}^{-iv_Pkt_0},
\end{align}
where $N_P$ is its initial pump photon number, and we have introduced the
Gaussian distribution
\begin{align}
\label{eq:Gaussian dist}
    \Phi(k) = \sqrt{\frac{v_P \tau}{\sqrt{\pi}}}\exp(-\frac{1}{2}\left[v_P\tau(k-K_P)\right]^2),
\end{align}
where $K_P$ and $v_P$ are the center wavenumber for the pulse and its group velocity in the actual channel, respectively, and $\tau$ is its duration. The distribution \eqref{eq:Gaussian dist} is normalized according to
\begin{align}
\int dk |\Phi(k)|^2 = 1.
\end{align}
A Fourier transform of Eq. \eqref{eq:initial pulse} for $x(k,t_0)$ yields the amplitude
as a function of position in the actual channel, i.e.,  $\int (dk/\sqrt{2\pi})x(k,t_0) {\rm e}^{ikz}$. We choose $t_0$ so that at the position $z = v_Pt_0$ where the pulse reaches its peak value in the actual channel is very far from the ring.

We now define the interaction picture ket as
\begin{align}
\label{eq:interaction picture ket}
    \ket{\psi(t)} \equiv {\rm e}^{iH_{\rm L}t/\hbar}\ket{\Psi(t)},
\end{align}
which satisfies the Schr\"odinger equation
\begin{align}
\label{eq:SE}
   i \hbar \frac{d\ket{\psi(t)}}{dt} = H_{\rm NL}(t) \ket{\psi(t)},
\end{align}
where $H_{\rm NL}(t) = {\rm e}^{iH_{\rm L} t/\hbar} H_{\rm NL}{\rm e}^{-iH_{\rm L} t/\hbar}$ is the interaction picture SPDC Hamiltonian. The initial ket in the interaction picture is independent of $t_0$. Putting $t=t_0$ into Eq. \eqref{eq:interaction picture ket} for $\ket{\psi(t)}$ and using Eq. \eqref{eq:initial state} for $\ket{\Psi(t_0)}$ we obtain
\begin{align}
\label{eq:input ket}
    \ket{\psi(t_0)} = \exp(\sqrt{N_P}\int dk\, \Phi(k) b^\dagger_{{\rm ac}Pk} - \rm H.c.)\ket{\rm vac}.
\end{align}
So as long as the pulse centered at the position $z=v_Pt_0$ in the input channel is far from the ring, the initial ket in the interaction picture \eqref{eq:input ket} is independent of $t_0$.  

We borrow the strategy of Yanagimoto et al. \cite{Yanagimoto2022} and look for a solution of Eq. \eqref{eq:SE} that separates the Gaussian and non-Gaussian dynamics as
\begin{align}
\label{eq:full ket}
    \ket{\psi(t)} = U(t)|\tilde{\psi}(t)\rangle,
\end{align}
where $U(t)$ is a Gaussian unitary that can be written as a product of displacement, squeezing, and rotation operators (see Appendix \ref{sec:U(t)}), and $|\tilde{\psi}(t)\rangle$ is a residual ket that describes the non-Gaussian dynamics. Putting Eq. \eqref{eq:full ket} for $\ket{\psi(t)}$ into Eq. \eqref{eq:SE} for the Schr\"odinger equation we obtain an equation that $|\tilde{\psi}(t)\rangle$ must satisfy
\begin{align}
\label{eq:non gaussian schrodinger equation}
   i \hbar \frac{d|\tilde{\psi}(t)\rangle}{dt} = H_{\rm eff}(t)|\tilde{\psi}(t)\rangle,
\end{align}
where at the initial time $t=t_0$ it is the vacuum state $|\tilde{\psi}(t_0)\rangle = \ket{\rm vac}$, and we have introduced the effective Hamiltonian
\begin{align}
\label{eq:Heff}
    H_{\rm eff}(t) &= U^\dagger(t)H_{\rm NL}(t)U(t) - i\hbar U^\dagger(t) \frac{dU(t)}{dt},
\end{align}
which is constructed to only contain non-Gaussian terms. To make $H_{\rm eff}(t)$ only contain non-Gaussian terms we choose a form for the Gaussian unitary $U(t)$ and adjust its squeezing, rotation, and displacement parameters so that the Gaussian terms in $H_{\rm eff}(t)$ cancel. 

In the two subsections below we derive the dynamical equations for the Gaussian parameters of $U(t)$ which make $H_{\rm eff}(t)$ purely non-Gaussian, and show how to solve Eq. \eqref{eq:non gaussian schrodinger equation} for the non-Gaussian ket $|\tilde{\psi}(t)\rangle$.

\subsection{Gaussian evolution}
In this subsection we derive the equations for the squeezing, rotation, and displacement parameters of the Gaussian unitary $U(t)$. The form of $U(t)$ we choose implements a displacement of the pump operator and a squeezing and rotation of the signal and idler operators, that is
\begin{align}
\label{eq:pump trans}
    U^\dagger(t) b_{P\mu_3}U(t)&=b_{P\mu_3} + [\bm x(t)]_{\mu_3},
    \\
    \label{eq:signal trans}
    U^\dagger(t) a_{S\mu_1}U(t)&=\sum_{\mu_2} [\bm V_{SS}(t)]_{\mu_1 \mu_2}a_{S\mu_2} \nonumber
    \\
    &+ \sum_{\mu_2}[\bm W_{SI}(t)]_{\mu_1\mu_2}a^\dagger_{I\mu_2},
    \\
    \label{eq:idler trans}
    U^\dagger(t) a_{I\mu_1}U(t)&=\sum_{\mu_2} [\bm V_{II}(t)]_{\mu_1 \mu_2}a_{I\mu_2} \nonumber
    \\
    &+\sum_{\mu_2} [\bm W_{IS}(t)]_{\mu_1\mu_2}a^\dagger_{S\mu_2},
\end{align}
where $\bm x(t)$ is a vector of pump displacement parameters and the matrices $\bm V_{SS}(t)$, $\bm V_{II}(t)$, $\bm W_{SI}(t)$, and $\bm W_{IS}(t)$ fully describe the characteristics of the squeezed state. These matrices can be written in terms of the squeezing and rotation parameters of a non-degenerate squeezing operator and the rotation operator, respectively, as we show in Appendix \ref{sec:U(t)}. Putting Eqs. \eqref{eq:pump trans}, \eqref{eq:signal trans}, and \eqref{eq:idler trans} in Eq. \eqref{eq:Heff} for $H_{\rm eff}(t)$ we show in Appendix \ref{sec:heff} that all Gaussian terms in $H_{\rm eff}(t)$ cancel if $\bm x(t)$, $\bm V_{SS}(t)$, $\bm V_{II}(t)$, $\bm W_{SI}(t)$, and $\bm W_{IS}(t)$ are solutions of the following differential equations:
\begin{align}
\label{eq:xdot}
i\frac{d\bm x}{dt}&= \bm \Delta(t),
\\
\label{eq:Vdot}
\begin{bmatrix}
    i\frac{d\bm V_{SS}}{dt} & 0 
    \\
    0 & i\frac{d\bm V_{II}}{dt}
\end{bmatrix}
  &= 
  \begin{bmatrix}
      0&\bm Z(t)
      \\
      \bm Z^{\rm T}(t) & 0
  \end{bmatrix}
    \begin{bmatrix}
      0&\bm W^*_{SI} (t)
      \\
      \bm W^*_{IS} (t) & 0
  \end{bmatrix},
  \\
  \label{eq:Wdot}
  \begin{bmatrix}
     0& i\frac{d\bm W_{SI}}{dt} 
    \\
    i\frac{d\bm W_{IS}}{dt}  & 0
\end{bmatrix}
  &= 
  \begin{bmatrix}
      0&\bm Z(t)
      \\
      \bm Z^{\rm T}(t) & 0
  \end{bmatrix}
    \begin{bmatrix}
      \bm V^*_{SS}(t)&0
      \\
      0& \bm V^*_{II}(t)
  \end{bmatrix},
\end{align}
where we have introduced the vector $\bm \Delta(t)$ as
\begin{align}
\label{eq:Delta}
    [\bm \Delta(t)]_{\mu_3}&=\sum_{\mu_1,\mu_2}[\Lambda^*(t)]_{\mu_1 \mu_2 \mu_3} [\bm W_{SI}(t) \bm V^{\rm T}_{II}(t)]_{\mu_1 \mu_2},
\end{align}
and the matrix $\bm Z(t)$ as
\begin{align}
    \label{eq:Z}
    [\bm Z(t)]_{\mu_1\mu_2} = \sum_{\mu_3}[\Lambda(t)]_{\mu_1 \mu_2 \mu_3}[\bm x(t)]_{\mu_3}.
\end{align}
Here $\bm \Delta(t)$ determines the rate at which the pump amplitude changes, which increases with increasing squeezing of the signal and idler fields. The matrix $\bm Z(t)$ characterizes
the squeezing of the signal and idler fields, which is driven
by the pump amplitude and enhanced by the resonant effects of the ring.

Since at our early time $t_0$ there
is no nonlinear interaction and the signal and idler are in the vacuum state, we have
$\bm W_{SI}(t_0) = \bm W_{IS}(t_0) = 0$, $\bm V_{SS}(t_0) = \bm V_{II}(t_0) = \bm I$, where $\bm I$ is the identity matrix. The initial coherent state for the pump \eqref{eq:input ket} has the amplitude $[\bm x(t_0)]_{\mu_3} = \sqrt{\Delta k}\sqrt{N_P} \Phi(k_3)$, where $\Phi(k_3)$ is given by Eq. \eqref{eq:Gaussian dist}. 
With these initial conditions we solve Eqs. \eqref{eq:xdot}, \eqref{eq:Vdot}, and \eqref{eq:Wdot} using a fourth-order Runge-Kutta method for the all the Gaussian parameters. 

Before moving on to the non-Gaussian evolution we discuss the Gaussian approximation, which assumes that the non-Gaussian ket is vacuum for all times
\begin{align}
\label{eq:Gaussian approx}
    |\tilde{\psi}(t)\rangle \approx \ket{\rm vac}.
\end{align}
In this case Eq. \eqref{eq:full ket} for the full ket is approximately given by the Gaussian unitary acting on the vacuum state. We define the Gaussian ket as
\begin{align}
\label{eq:Gaussian ket}
    \ket{\psi_G(t)} \equiv U(t) \ket{\rm vac}.
\end{align}
Using the Gaussian ket, the number of pump photons at all times is given by the inner product 
\begin{align}
    N_P(t) \equiv \bra{\psi_G(t)} \sum_{\mu_3}b^\dagger_{P\mu_3} b_{P\mu_3} \ket{\psi_G(t)} = [\bm x^*(t)]^{\rm T} \bm x(t),
\end{align}
where the amount of pump depletion is defined as
\begin{align}
\label{eq:pump depletion}
    d(t) \equiv 1 - \frac{N_P(t)}{N_P},
\end{align}
and the second moments are defined as
\begin{align}
\label{eq:M}
   [\bm M_{SI}(t)]_{\mu_1 \mu_2}&\equiv  \bra{\psi_G(t)} a_{S\mu_1}a_{I\mu_2} \ket{\psi_G(t)} \nonumber
   \\
   &= [\bm W_{SI}(t) \bm V^{\rm T}_{II}(t)]_{\mu_1\mu_2},
    \\
    \label{eq:NS}
    [\bm N_{SS}(t)]_{\mu_1 \mu_2}&\equiv  \bra{\psi_G(t)} a^\dagger_{S\mu_1}a_{S\mu_2} \ket{\psi_G(t)} \nonumber
    \\
    &= [\bm W^*_{SI}(t) \bm W^{\rm T}_{SI}(t)]_{\mu_1\mu_2},
    \\
    \label{eq:NI}
    [\bm N_{II}(t)]_{\mu_1 \mu_2}&\equiv  \bra{\psi_G(t)} a^\dagger_{I\mu_1}a_{I\mu_2} \ket{\psi_G(t)} \nonumber
    \\
    &= [\bm W^*_{IS}(t) \bm W^{\rm T}_{IS}(t)]_{\mu_1\mu_2}.
\end{align}
These moments can be partitioned into into $2\times 2$ block matrices involving the four different combinations of output channels
\begin{align}
\label{eq:M block}
    \bm M_{SI}(t) &= 
    \begin{bmatrix}
        \bm M^{\rm ac\, ac}_{SI}(t) & \bm M^{\rm ac\, ph}_{SI}(t)
        \\
        \bm M^{\rm ph\, ac}_{SI}(t) & \bm M^{\rm ph\, ph}_{SI}(t)
    \end{bmatrix},
    \\
    \label{eq:N block}
    \bm N_{JJ}(t) &= 
    \begin{bmatrix}
        \bm N^{\rm ac\, ac}_{JJ}(t) & \bm N^{\rm ac\, ph}_{JJ}(t)
        \\
        \bm N^{\rm ph\, ac}_{JJ}(t) & \bm N^{\rm ph\, ph}_{JJ}(t)
    \end{bmatrix},
\end{align}
for $J\in\{S,I\}$. The block matrices labeled by $(\rm ac\, ac)$ and $(\rm ph\, ph)$ account for cases where both signal and idler are in the actual output and phantom output channels, respectively, while those labeled by $(\rm ac\, ph)$ and $(\rm ph\, ac)$ account for cases where the signal and idler pairs are split across the two output channels. 

Given $\bm{x}(t)$, $\bm{M}_{SI}(t)$, $\bm{N}_{SS}(t)$, and $\bm{N}_{II}(t)$, we obtain a complete description of the Gaussian dynamics. We emphasize that this solution fully describes non-degenerate squeezing of the signal and idler fields for arbitrary pump energies, including pump depletion and scattering loss.

\subsection{Non-Gaussian evolution}
 We now turn our attention to the non-Gaussian dynamics. Given that the Gaussian parameters $\bm x(t)$, $\bm V_{SS}(t)$, $\bm V_{II}(t)$, $\bm W_{SI}(t)$, and $\bm W_{IS}(t)$ are solutions to the differential Eqs. \eqref{eq:xdot}--\eqref{eq:Wdot} this guarantees that all the Gaussian terms in Eq. \eqref{eq:Heff} for $H_{\rm eff}(t)$ cancel, leaving only the non-Gaussian terms  (see Appendix \ref{sec:heff})
\begin{align}
\label{eq:Heff NG}
    H_{\rm eff}(t) &= \sum_{\mu_1,\mu_2,\mu_3,\mu_1',\mu_2'}\hbar [\Lambda^*(t)]_{\mu_1 \mu_2 \mu_3}\nonumber
    \\
    &\times \big([\bm V_{SS}(t)]_{\mu_1 \mu_1'}[\bm V_{II}(t)]_{\mu_2 \mu_2'} a_{I\mu_2'} a_{S\mu_1'}b^\dagger_{P\mu_3}  \nonumber
    \\
    &+[\bm W_{SI}(t)]_{\mu_1 \mu_1'}[\bm V_{II}(t)]_{\mu_2 \mu_2'} a^\dagger_{I\mu_1'} a_{I\mu_2'}b^\dagger_{P\mu_3} \nonumber
      \\
    &+[\bm V_{SS}(t)]_{\mu_1 \mu_1'}[\bm W_{IS}(t)]_{\mu_2 \mu_2'} a^\dagger_{S\mu_2'} a_{S\mu_1'}b^\dagger_{P\mu_3}  \nonumber
      \\
    & +[\bm W_{SI}(t)]_{\mu_1 \mu_1'}[\bm W_{IS}(t)]_{\mu_2 \mu_2'} a^\dagger_{I\mu_1'} a^\dagger_{S\mu_2'}b^\dagger_{P\mu_3} \big)\nonumber
    \\
    &+{\rm H.c.}.
\end{align}
Given that Eq. \eqref{eq:Heff NG} for $H_{\rm eff}(t)$ only contains non-Gaussian terms, Eq. \eqref{eq:non gaussian schrodinger equation} for the residual Schr\"odinger equation yields a $|\tilde{\psi}(t)\rangle$  that describes only the non-Gaussian features of the light.

Before obtaining a numerical solution for the non-Gaussian ket $|\tilde{\psi}(t)\rangle$ using Eq. \eqref{eq:Heff NG} for $H_{\rm eff}(t)$ we seek a perturbative Dyson series solution under the assumption that the effect of $H_{\rm eff}(t)$ acting on the vacuum is small. Only keeping up to the first-order term in the Dyson series, the ket can be approximately written as
\begin{align}
\label{eq:perturb psi tilde}
    |\tilde{\psi}(t)\rangle \approx \ket{{\rm vac}} + |\tilde{\psi}_1(t)\rangle,
\end{align}
where we have defined the first-order correction to the ket as
\begin{align}
\label{eq:psi tilde 1}
    |\tilde{\psi}_1(t)\rangle&=-i\sum_{\mu'_1 ,\mu'_2,\mu_3} [\mathcal{O}(t)]_{\mu'_1 \mu'_2 \mu_3}a^\dagger_{I\mu_1'} a^\dagger_{S\mu_2'}b^\dagger_{P\mu_3}\ket{{\rm vac}},
\end{align}
where
\begin{align}
    [\mathcal{O}(t)]_{\mu'_1\mu'_2 \mu_3}&=\sum_{\mu_1, \mu_2 }\int_{t_0}^t dt' [\Lambda^*(t')]_{\mu_1 \mu_2 \mu_3}\nonumber
    \\
    &\times [\bm W_{SI}(t')]_{\mu_1 \mu_1'}[\bm W_{IS}(t')]_{\mu_2 \mu_2'}. 
\end{align}
The perturbative solution in Eq. \eqref{eq:perturb psi tilde} is valid as long as the norm
of the correction is much less than unity for all $t$,
\begin{align}
\label{eq:overlap}
    \langle \tilde{\psi}_1(t)|\tilde{\psi}_1(t)\rangle \ll 1\,\,\,\rightarrow \,\,\,\sum_{\mu'_1 ,\mu'_2,\mu_3}\left|[\mathcal{O}(t)]_{\mu'_1\mu'_2\mu_3}\right|^2\ll1.
\end{align}
 The Gaussian approximation would be exact were the norm
 of the correction
 zero; a nonzero norm
 indicates
 a deviation from the Gaussian regime. As the value of the norm becomes significant,
 the perturbative non-Gaussian solution \eqref{eq:perturb psi tilde} should not be trusted either. This occurs when the generation of signal and idler photons is very efficient, as we 
 show in Sec. \ref{sec:results}.

When
$\langle \tilde{\psi}_1(t)|\tilde{\psi}_1(t)\rangle$ 
becomes non-negligible, we must move beyond the perturbative solution \eqref{eq:perturb psi tilde} and look for a numerical solution of Eq. \eqref{eq:non gaussian schrodinger equation} for $|\tilde{\psi}(t) \rangle$. However, it is numerically impractical to obtain $|\tilde{\psi}(t) \rangle$ using Eq. \eqref{eq:Heff NG}, where 
$H_{\rm eff}(t)$ is written in terms of asymptotic field operators. A more practical method 
is to express $H_{\rm eff}(t)$ in a basis of supermodes that can be truncated to a few that are dominant.
This strategy has been investigated in previous work \cite{Yanagimoto2022} in the context of degenerate squeezing in PPLN; in what follows we adapt this strategy to identify the dominant supermodes in our microring system.

We identify the dominant supermodes based on the heuristic assumption that non-Gaussian features are concentrated in supermodes with substantial squeezing. The information about the squeezed state is contained in the matrices $\bm V_{SS}(t)$, $\bm V_{II}(t)$, $\bm W_{SI}(t)$, and $\bm W_{IS}(t)$. To extract the modes that are most squeezed we perform the following joint singular value decomposition (SVD) of the matrices \cite{houde2024matrix}
\begin{align}
\label{eq:joint svd}
    \bm V_{SS}(t) &= \bm F_S(t) \cosh[\bm r(t)] \bm G_S(t), \nonumber
        \\
    \bm V_{II}(t) &= \bm F^*_I(t) \cosh[\bm r(t)] \bm G^*_I(t), \nonumber
    \\
     \bm W_{SI}(t) &= \bm F_S(t) \sinh[\bm r(t)] \bm G_I(t), \nonumber
    \\
    \bm W_{IS}(t) &= \bm F^*_I(t) \sinh[\bm r(t)] \bm G^*_S(t), 
\end{align}
where $\bm F_S(t)$, $\bm F_I(t)$, $\bm G_S(t)$, and $\bm G_I(t)$ are unitary matrices, $\bm r(t) = {\rm diag}(r_1(t), r_2(t), \ldots)$ is a diagonal matrix, where $r_u(t) \ge 0$ is the squeezing amplitude for supermode $u$. Here we can write $\cosh[\bm r(t)] = {\rm diag}(\cosh[r_1(t)], \cosh[r_2(t)], \ldots)$ and $\sinh[\bm r(t)] = {\rm diag}(\sinh[r_1(t)], \sinh[r_2(t)], \ldots)$. Putting Eq. \eqref{eq:joint svd} for the joint SVD into Eq. \eqref{eq:Heff NG} for $H_{\rm eff}(t)$ we obtain
\begin{widetext}
\begin{align}
\label{eq:Heff NG 2}
    H_{\rm eff}(t) &= \sum_{u,u'}\,\sum_{\mu_3,\mu_1',\mu_2'}\hbar [\bm L(t)]_{\mu_3 u u'}b^\dagger_{P\mu_3}\Big(\cosh[r_u(t)] \cosh[r_{u'}(t)] [\bm G^*_I(t)]_{u' \mu_2'} a_{I\mu_2'} [\bm G_S(t)]_{u \mu_1'}a_{S\mu_1'}  \nonumber
    \\
    &+\sinh[r_u(t)] \cosh[r_{u'}(t)] [\bm G_I(t)]_{u' \mu_1'} a^\dagger_{I\mu_1'} [\bm G^*_I(t)]_{u \mu_2'}a_{I\mu_2'}+\cosh[r_u(t)] \sinh[r_{u'}(t)] [\bm G^*_S(t)]_{u' \mu_2'} a^\dagger_{S\mu_2'} [\bm G_S(t)]_{u \mu_1'}a_{S\mu_1'} \nonumber
      \\
&+\sinh[r_u(t)] \sinh[r_{u'}(t)] [\bm G_I(t)]_{u' \mu_1'} a^\dagger_{I\mu_1'} [\bm G^*_S(t)]_{u \mu_2'}a^\dagger_{S\mu_2'}\Big)+{\rm H.c.},
\end{align}
\end{widetext}
where we have introduced the tensor
\begin{align}
\label{eq:L tensor}
    [\bm L(t)]_{\mu_3 uu'}&= \sum_{\mu_1,\mu_2}[\Lambda^*(t)]_{\mu_1 \mu_2 \mu_3}[\bm F_S(t)]_{\mu_1 u}[\bm F^*_I(t)]_{\mu_2 u'}.
\end{align}
With
Eq. \eqref{eq:Heff NG 2} in hand it is natural to define the supermode annihilation operators for the signal and idler as
\begin{align}
\label{eq:AS}
    A_{S u}(t) &= \sum_{\mu'_1}[\bm G_S(t)]_{u\mu'_1} a_{S\mu'_1},
    \\
    \label{eq:AI}
    A_{I u}(t) &= \sum_{\mu'_2}[\bm G^*_I(t)]_{u\mu'_2} a_{I\mu'_2},
\end{align}
where $\bm G_S(t)$ and $\bm G_I(t)$ are the unitary matrices from the joint SVD \eqref{eq:joint svd}. We can write Eq. \eqref{eq:Heff NG 2} for $H_{\rm eff}(t)$ in terms of the signal and idler supermode operators as
\begin{align}
\label{eq:Heff NG 3}
    H_{\rm eff}(t) &= \sum_{u,u',\mu_3}\hbar [\bm L(t)]_{\mu_3 u u'}b^\dagger_{P\mu_3}\nonumber
    \\
    &\times \Big(\cosh[r_u(t)] \cosh[r_{u'}(t)] A_{Iu'}(t)A_{Su}(t)  \nonumber
    \\
    &+\sinh[r_u(t)] \cosh[r_{u'}(t)] A^\dagger_{Iu'}(t)A_{Iu}(t)\nonumber
    \\
    &+\cosh[r_u(t)] \sinh[r_{u'}(t)] A^\dagger_{Su'}(t)A_{Su}(t)\nonumber
      \\
&+\sinh[r_u(t)] \sinh[r_{u'}(t)] A^\dagger_{Iu'}(t)A^\dagger_{Su}(t)\Big)+{\rm H.c.}.
\end{align}

Considering Eq. \eqref{eq:Heff NG 3} for $H_{\rm eff}(t)$, the tensor element $[\bm L(t)]_{\mu_3 u u'}$ describes the coupling between the pump supermode labeled by $\mu_3$ and the pair of signal and idler supermodes labeled by $(u, u')$. In general every element of $\bm L(t)$ is nonzero, which implies that each pump mode is coupled with some strength to every signal-idler pair. But for efficient simulation we aim to identify a single pump supermode that couples predominantly to a subset of signal and idler supermodes.  To try to isolate the pump from the signal and idler, we perform an SVD of $\bm L(t)$, reorganizing
Eq. \eqref{eq:L tensor} for the tensor $\bm L(t)$ into a matrix by grouping together the signal and idler supermode indices as $(u,u')$. We perform the following SVD on the reshaped $\bm L(t)$ matrix
\begin{align}
\label{eq:L svd}
    [\bm L(t)]_{\mu_3 (u,u')} = \sum_{\lambda}[\bm X(t)]_{\mu_3 \lambda}[\bm D(t)]_{\lambda \lambda} [\bm Q^\dagger(t)]_{\lambda (u,u')},
\end{align}
where $\bm X(t)$ is the square unitary matrix that results, $\bm Q(t)$ is a non-square semi-unitary matrix, with $\bm Q^\dagger(t) \bm Q(t) = \bm I$ but $\bm Q(t) \bm Q^\dagger(t) \neq \bm I$, and $\bm D(t) = {\rm diag}(D_1(t), D_2(t), \ldots)$ is a diagonal matrix with $D_{\lambda}(t)\ge 0$. Putting Eq. \eqref{eq:L svd} for $\bm L(t)$ into Eq. \eqref{eq:Heff NG 3} for $H_{\rm eff}(t)$ we obtain
\begin{align}
\label{eq:Heff NG 4}
    H_{\rm eff}(t) &= \sum_{u,u',\lambda}\hbar D_{\lambda}(t)[\bm Q^\dagger(t)]_{\lambda(u,u')} B^\dagger_{P\lambda}(t)\nonumber
    \\
    &\times \Big(\cosh[r_u(t)] \cosh[r_{u'}(t)] A_{Iu'}(t)A_{Su}(t)  \nonumber
    \\
    &+\sinh[r_u(t)] \cosh[r_{u'}(t)] A^\dagger_{Iu'}(t)A_{Iu}(t)\nonumber
    \\
    &+\cosh[r_u(t)] \sinh[r_{u'}(t)] A^\dagger_{Su'}(t)A_{Su}(t)\nonumber
      \\
&+\sinh[r_u(t)] \sinh[r_{u'}(t)] A^\dagger_{Iu'}(t)A^\dagger_{Su}(t)\Big)+{\rm H.c.},
\end{align}
where we have introduced the pump supermode annihilation operators as
\begin{align}
\label{eq:BP}
    B_{P\lambda}(t) = \sum_{\mu_3}[\bm X^*(t)]_{\mu_3 \lambda}b_{P\mu_3}, 
\end{align}
where $\bm X(t)$ is the square unitary matrix from the SVD \eqref{eq:L svd}.

Eq. \eqref{eq:Heff NG 4} is our
final result for $H_{\rm eff}(t)$, written in terms of the supermodes. Each pump supermode labeled by $\lambda$ is coupled with every pair of signal and idler supermodes labeled by $(u,u')$,  with the strength of each connection weighted by $D_{\lambda}(t)[\bm Q^\dagger(t)]_{\lambda(u,u')}$. The benefit of having the form \eqref{eq:Heff NG 4} is that the pump singular values $D_\lambda$ are organized in decreasing magnitude, so we obtain an ordering of the coupling strength between a pump mode and each signal-idler pair that we can exploit. For example, using realistic parameters we demonstrate in Sec. \ref{sec:results} 
that all $D_{\lambda}(t)$ are  virtually zero except for one. 
So effectively 
a single pump supermode 
couples to every signal-idler pair. 
Additionally, 
we find that keeping only a few signal and idler supermodes in $H_{\rm eff}(t)$ is sufficient to capture the total number of photons in the non-Gaussian ket. These observations justify truncating Eq. \eqref{eq:Heff NG 4} for $H_{\rm eff}(t)$ to a small number of supermodes. In the truncated supermode basis we can perform efficient numerical simulations to obtain the non-Gaussian ket.

 We stress that Eqs. \eqref{eq:AS} and \eqref{eq:AI} for the signal and idler supermode operators annihilate photons in supermodes that have components in both the actual and the phantom channels.
 So Eq. \eqref{eq:Heff NG 4} for $H_{\rm eff}(t)$ describes the evolution of the non-Gaussian ket in the entire system.

With Eq. \eqref{eq:Heff NG 4} for $H_{\rm eff}(t)$ 
we are now in a position to solve Eq. \eqref{eq:non gaussian schrodinger equation} for the Schr\"odinger equation for the non-Gaussian ket $|\tilde{\psi}(t)\rangle$. To benefit from the supermode basis we represent $|\tilde{\psi}(t)\rangle$ in the basis of time-dependent supermode Fock states given by
\begin{align}
\label{eq:signal fock ket}
    \ket{n_{Su}}_t &= \frac{1}{\sqrt{n_{Su}!}}[A^\dagger_{Su}(t)]^{n_{Su}}\ket{0},
    \\
\label{eq:idler fock ket}
    \ket{n_{Iu}}_t &= \frac{1}{\sqrt{n_{Iu}!}}[A^\dagger_{Iu}(t)]^{n_{Iu}}\ket{0},
    \\
\label{eq:pump fock ket}
    \ket{n_{Pu}}_t &= \frac{1}{\sqrt{n_{Pu}!}}[B^\dagger_{Pu}(t)]^{n_{Pu}}\ket{0},
\end{align}
where at each time $n_{Su}$, $n_{Iu}$, and $n_{Pu}$ are the number of photons in the signal, idler, and pump supermode $u$ respectively. We consider a truncated basis consisting of $M$ signal kets, $M$ idler kets, and $L$ pump kets, where each ket contains at most $N$ photons. We choose the same number of signal and idler kets, but this can be readily generalized. The non-Gaussian ket can be written in this truncated basis as
\begin{align}
\label{eq:psi tilde numbers}
|\tilde{\psi}(t)\rangle   &= \sum_{n_{S1},\ldots, n_{SM} = 0}^N 
\,\,\,\sum_{n_{I1},\ldots, n_{IM} = 0}^N 
\,\,\,\sum_{n_{P1},\ldots, n_{PL} = 0}^N \nonumber
\\
&\times c_{n_{S1}\ldots n_{SM}n_{I1}\ldots n_{IM}n_{P1}\ldots n_{PL}}(t) \nonumber
\\
&\times \ket{n_{S1}\ldots n_{SM}}_t\ket{n_{I1}\ldots n_{IM}}_t\ket{n_{P1}\ldots n_{PL}}_t,
\end{align}
where $c_{n_{S1}\ldots n_{SM}n_{I1}\ldots n_{IM}n_{P1}\ldots n_{PL}}(t)$ are the coefficients that we 
determine. Supermode basis kets outside the truncated basis are assumed to be in the vacuum state, and are not included in the expansion \eqref{eq:psi tilde numbers}. Taking the time derivative of Eq. \eqref{eq:psi tilde numbers} for $|\tilde{\psi}(t)\rangle$ we obtain a linear Hamiltonian $\tilde{H}_{\rm L}(t)$ given below that captures the time dependence of the supermode basis. Putting Eq. \eqref{eq:psi tilde numbers} into Eq. \eqref{eq:non gaussian schrodinger equation} for $i\hbar\, d|\tilde{\psi}(t)\rangle/dt$, we show in Appendix \ref{sec:dpsidt} that we can write
\begin{align}
\label{eq:psi tilde numerical}
    i\hbar \frac{\partial |\tilde{\psi}(t)\rangle}{\partial t} = \left(H_{\rm eff}(t) + \tilde{H}_{\rm L}(t)\right)|\tilde{\psi}(t)\rangle,
\end{align}
where $H_{\rm eff}(t)$ is given by Eq. \eqref{eq:Heff NG 4} and we have defined
\begin{align}
\label{eq:partial psi tilde}
\frac{\partial |\tilde{\psi}(t)\rangle}{\partial t}    &= \sum_{n_{S1},\ldots, n_{SM} = 0}^N \sum_{n_{I1},\ldots, n_{IM} = 0}^N  \sum_{n_{P1},\ldots, n_{PL} = 0}^N \nonumber
\\
&\times \frac{\partial }{\partial t}\left[ c_{n_{S1}\ldots n_{SM}n_{I1}\ldots n_{IM}n_{P1}\ldots n_{PL}}(t)\right] \nonumber
\\
&\times \ket{n_{S1}\ldots n_{SM}}_t\ket{n_{I1}\ldots n_{IM}}_t\ket{n_{P1}\ldots n_{PL}}_t.
\end{align}
The linear Hamiltonian $\tilde{H}_{\rm L}(t)$, which arises because the supermodes themselves are time dependent, is given by
\begin{align}
\label{eq:HL supermode}
    \tilde{H}_L(t) &= -i\hbar\left[\frac{d \bm G^*_S(t)}{dt}\bm G^{\rm T}_{S}(t)\right]_{uu'}A^\dagger_{Su'}(t)A_{Su}(t) \nonumber
    \\
    &-i\hbar\left[\frac{d \bm G_I(t)}{dt}\bm G^{\dagger}_{I}(t)\right]_{uu'}A^\dagger_{Iu'}(t)A_{Iu}(t) \nonumber
    \\
    &-i\hbar \left[\frac{d \bm X^{\rm T} (t)}{dt}\bm X^*(t)\right]_{uu'}B^\dagger_{Pu'}(t)B_{Pu}(t),
\end{align}
 where $\bm G_{S}(t)$ and $\bm G_{I}(t)$, and $\bm X(t)$ are the unitary matrices from the SVDs in Eqs. \eqref{eq:joint svd}, and \eqref{eq:L svd} respectively. 

We numerically solve Eq. \eqref{eq:psi tilde numerical} for the coefficients of $|\tilde{\psi}(t)\rangle$ by evolving the state over short time steps $\Delta t$, during which the time dependence of the Hamiltonian can be neglected. This approach allows us to efficiently obtain the non-Gaussian ket for a number of supermodes on the order of 1 ($M\approx L \approx 1$) and a Fock space cutoff for the non-Gaussian ket on the order of 10 ($N\approx 10$).

\section{Results}
\label{sec:results}
In this section we present our calculations.
We model a squeezing measurement in the output channel and calculate the conversion efficiency of the signal and idler,
compute the Wigner function of the non-Gaussian ket, and
discuss how the non-Gaussian features of the full ket could be observed in the actual output channel.The
non-Gaussian ket and the signal and idler kets
start in the vacuum state, and we choose the initial pump field in the actual input channel state to take the form of the Gaussian pulse in Eqs. \eqref{eq:initial pulse} and \eqref{eq:Gaussian dist}.  We assume that the pump is always critically coupled to the ring, $\eta_P = 0.5$, but allow its duration and intrinsic quality factor to vary.

\subsection{Gaussian results}

 We begin with the calculation of a homodyne measurement of the squeezed state in the actual output channel; details are presented 
 in Appendix \ref{sec:lo shapes}. We evaluate our prediction for the measurement quadrature noise (Eq. \eqref{eq:Xnoise})
 using the reduced moments $\bm M^{\rm ac\,ac}_{SI}(t_1)$, $\bm N^{\rm ac\,ac}_{SS}(t_1)$, and $\bm N^{\rm ac\,ac}_{II}(t_1)$, where $t_1 = 30/\bar{\Gamma}_P$ is a late time when the pump pulse has left the ring and the generated light is far from the ring. The reduced moments are obtained numerically from the coupled Eqs. \eqref{eq:xdot}, \eqref{eq:Vdot}, and \eqref{eq:Wdot} for the Gaussian parameters. The quadrature noise is squeezed when it reaches below the vacuum noise, which we normalize to 1, and the minimum in the quadrature noise is achieved by using the optimal local oscillator (LO) shapes that are obtained by a Lagrange multiplier method, as discussed in Appendix \ref{sec:lo shapes}.

\begin{figure}[htbp]
    \centering
    \includegraphics[scale=0.5
    ]{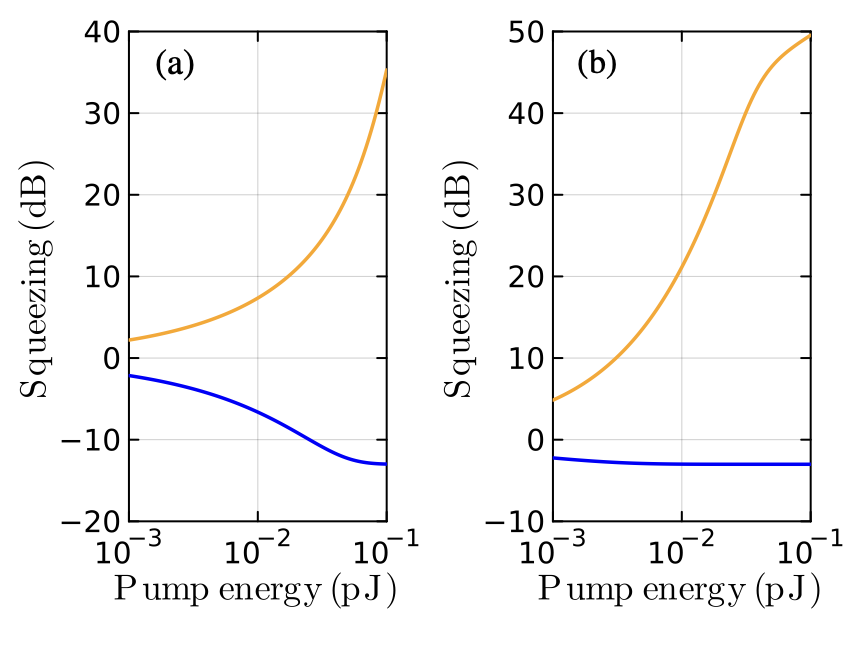}
    \caption{Squeezing (blue) and anti-squeezing (orange) in the actual output channel versus input pump energy for (a) $\eta_S=\eta_I=0.95$ and (b) $\eta_S=\eta_I=0.5$. The squeezing measurement is done at a late time $t_1 = 30/\bar{\Gamma}_P$, where $\bar{\Gamma}_P = 2\pi \times 3.84\,{\rm GHz}$,  when the pump pulse has left the ring. We have taken
    a pump pulse duration of $0.5\,{\rm ns}$, and used $Q_{{\rm int}, S} = Q_{{\rm int}, I} = 10^6$.}
    \label{fig:squeezing}
\end{figure}

In Fig. \ref{fig:squeezing}(a) the squeezing (blue) and anti-squeezing (orange) in the actual channel are shown as functions of the input pump energy for overcoupling of the signal and idler, $\eta_S = \eta_I = 0.95$. The asymmetric squeezing and anti-squeezing is due to the mixed-state of the actual output channel that arises from tracing over the phantom channel in Eq. \eqref{eq:Xnoise} for the quadrature noise. The squeezing saturates at approximately $-12.99\,{\rm dB}$ for an energy of $0.1\,{\rm pJ}$. We show in Appendix \ref{sec:lo shapes} that the limiting squeezing level when $\eta_S = \eta_I$ is approximately given by $\langle X^2\rangle_{\rm min} \approx 1-(\eta_S+\eta_I)/2$, indicating that the minimum noise is limited by the escape efficiency into the actual channel.  So for $\eta_S = \eta_I = 0.95$ the analytical minimum noise is given by $\langle X^2\rangle_{\rm min} \approx 0.05$, which is about $-13.01\,{\rm dB}$. In Fig. \ref{fig:squeezing}(b) the squeezing (blue) and anti-squeezing (orange) are shown for critical coupling, $\eta_S = \eta_I = 0.5$. The squeezing saturates at approximately $-3.01\,{\rm dB}$ for an energy of $0.1\,{\rm pJ}$, which is virtually identical to the analytical result for the minimum noise $\langle X^2\rangle_{\rm min} \approx 1/2$.

\begin{figure}[htbp]
    \centering
    \includegraphics[scale=0.133]{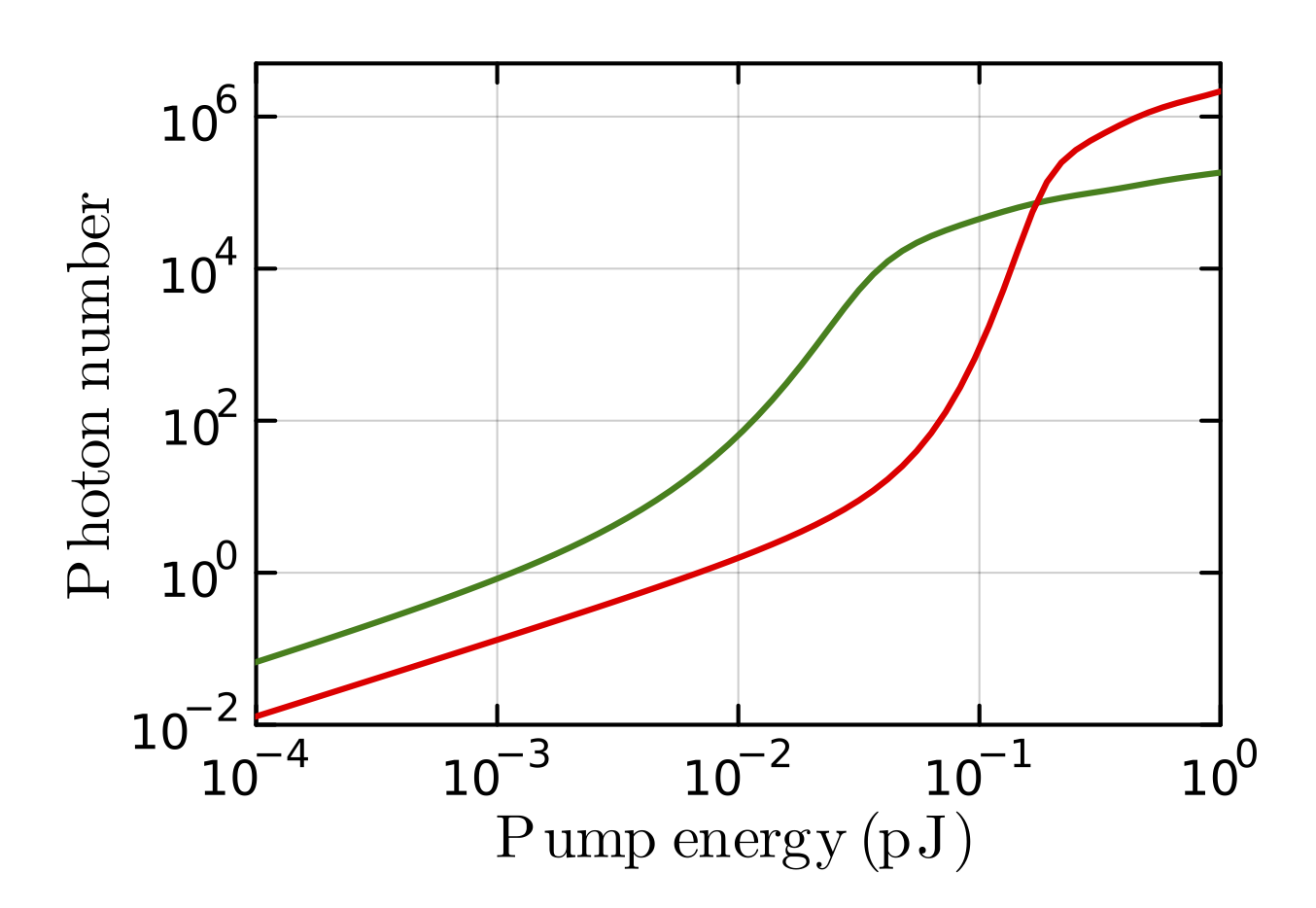}
    \caption{Signal photon number (identical to the idler photon number) versus the initial pump energy for $\eta_S = \eta_I = 0.5$ (green) and $\eta_S = \eta_I = 0.95$ (red), evaluated 
    at $t_1 = 30/\bar{\Gamma}_P$ with
    $\bar{\Gamma}_P = 2\pi \times 3.84\,{\rm GHz}$. We have taken 
    a pump pulse duration of $0.5\,{\rm ns}$, and used  $Q_{{\rm int}, S} = Q_{{\rm int}, I} = 10^6$.}
    \label{fig:number}
\end{figure}

We now calculate the number of generated signal photons, which is identical to the number of generated idler photons. The photon number is given by ${\rm tr}[\bm N_{SS}(t_1)] = {\rm tr}[\bm N_{II}(t_1)]$, where the photons in the phantom channel are included. In Fig. \ref{fig:number} the photon number versus initial pump energy for for critical coupling (green) and overcoupling (red) is shown. We evaluate the photon number at a very late time when the pump pulse has left the ring. For low pump energy the photon number is proportional to the energy, which is indicative of the photon-pair regime \cite{Fontaine2025}. Increasing the pump energy causes the photon number to increase exponentially due to the creation of a squeezed state, and then a gradual saturation of the photon number indicates the onset of the pump depletion regime \cite{Florez2020}.

\begin{figure}[htbp]
    \centering
    \includegraphics[scale=0.55]{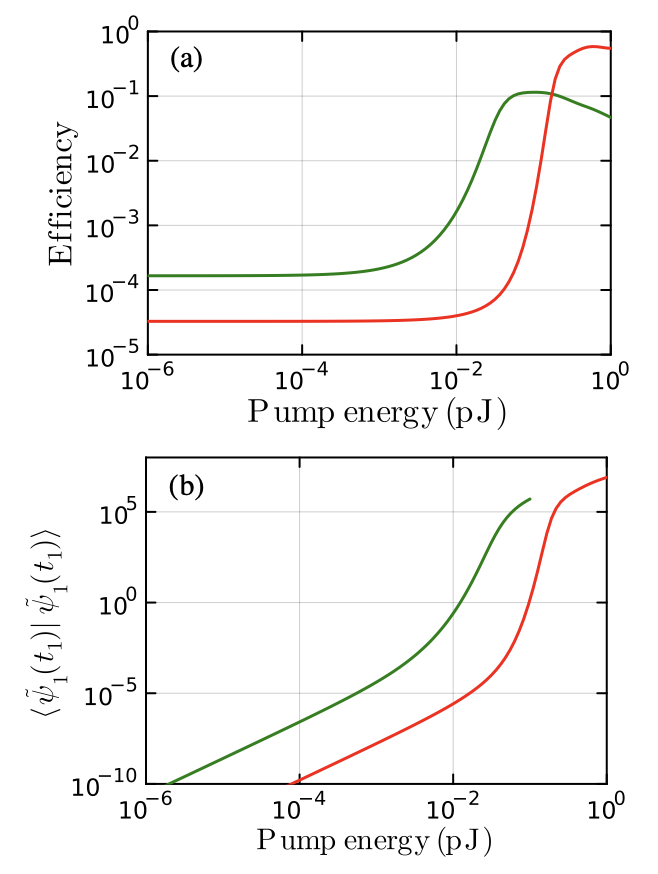}
    \caption{(a) Conversion efficiency and (b) overlap of the perturbative non-Gaussian ket versus the initial pump energy for $\eta_S = \eta_I = 0.5$ (green) and $\eta_S = \eta_I = 0.95$ (red), evaluated 
    at $t_1 = 30/\bar{\Gamma}_P$ with
    $\bar{\Gamma}_P = 2\pi \times 3.84\,{\rm GHz}$. We have taken
    a pump pulse duration of $0.5\,{\rm ns}$, and used $Q_{{\rm int}, S} = Q_{{\rm int}, I} = 10^6$.}
    \label{fig:ceff}
\end{figure}

The amount of pump depletion (see Eq. \eqref{eq:pump depletion}) is equal to the conversion efficiency of pump photons into signal and idler photons. The conversion efficiency is defined as the ratio of the number of generated signal photons after the pump pulse has left the ring to the initial number of pump photons. 
In Fig. \ref{fig:ceff}(a) the conversion efficiency versus the initial pump energy is shown. In the photon-pair regime the 
efficiency is independent of the pump energy, 
beyond that regime it first increases exponentially with increasing pump energy, and then starts to decrease in the pump depletion regime. The decrease in the efficiency is due to the gradual saturation of the photon number as
the initial pump photon number continues to increase.  For a pump energy of $1\,{\rm pJ}$ the efficiencies are about $4.6\%$ (green) and $54.8\%$ (red) for critical coupling and overcoupling, respectively. For low pump energies and critical coupling we obtain an efficiency independent of energy of $1.66\times 10^{-4}$, which is the same order of magnitude observed in similar InGaP microrings \cite{Akin2024}. 

To estimate the validity of the Gaussian approximation in the pump depletion regime we calculate the norm
of the perturbative non-Gaussian ket $|\tilde{\psi}_1(t_1)\rangle$ (see Eq. \eqref{eq:overlap}). In Fig. \ref{fig:ceff}(b) the norm 
versus the initial pump energy is shown. The norm
is approximately unity $\langle\tilde{\psi}_1(t_1)|\tilde{\psi}_1(t_1)\rangle \approx 1$ at $0.013\,{\rm pJ}$ for critical coupling (green)  and at $0.1\,{\rm pJ}$ for overcoupling (red), indicating that the Gaussian approximation is no longer valid at these respective energies.

\begin{figure}[htbp]
    \centering
    \includegraphics[scale=0.6]{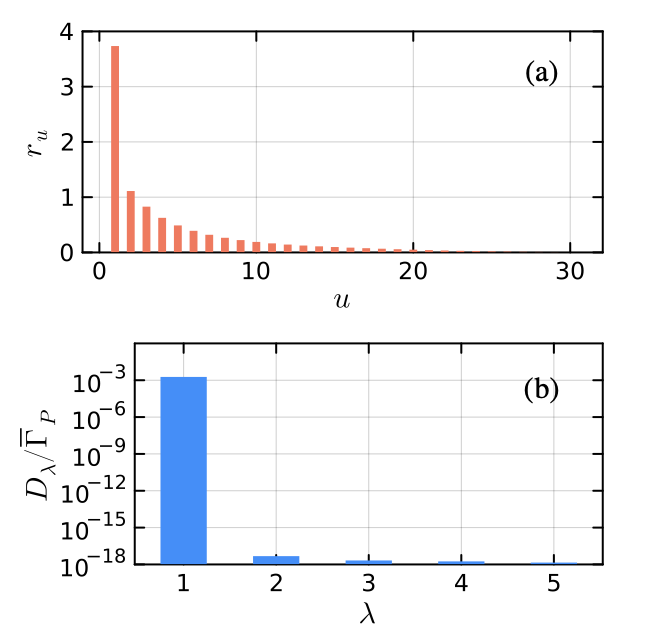}
    \caption{(a)  Largest 30 singular values associated with the signal and idler (see Eq. \eqref{eq:joint svd}) and (b) largest 5 singular values associated with the pump (see Eq. \eqref{eq:L svd}), for $\eta_S = \eta_I = 0.95$ and a pump energy of $0.09\,{\rm pJ}$, evaluated
    at $t_1 = 30/\bar{\Gamma}_P$ with
    $\bar{\Gamma}_P = 2\pi \times 3.84\,{\rm GHz}$. We have taken
    a pump pulse duration of $0.5\,{\rm ns}$, and used $Q_{{\rm int}, S} = Q_{{\rm int}, I} = 10^6$. }
    \label{fig:singular values}
\end{figure}

\subsection{Non-Gaussian results}

Having identified the pump energy regimes in
which the Gaussian approximation starts to fail, we now turn our attention to the non-Gaussian ket. 

We begin by considering an escape efficiency of $\eta_S = \eta_I = 0.95$ and a pump energy of $0.09\,{\rm pJ}$ where  $\langle\tilde{\psi}_1(t_1)|\tilde{\psi}_1(t_1)\rangle \approx 1$ and the onset of non-Gaussianity is expected. In Fig. \ref{fig:singular values}(a) the leading 30 singular values associated with the signal and idler supermodes squeezing coefficients $r_u$ are shown. The first singular value is approximately $r_1 = 3.7$, and $r_{10}$, $r_{20}$ and $r_{30}$ are approximately $0.19$, $0.05$, and $0.006$ respectively. This indicates that the effective Hamiltonian and the resulting evolution of the non-Gaussian ket are dominated by the first signal and idler supermodes. In Fig. \ref{fig:singular values}(b) the largest 5 singular values of the pump supermodes $D_\lambda$ are shown. The singular values other than the first one are virtually zero, indicating that keeping only the first pump supermode in the effective Hamiltonian is sufficient to model the evolution of the non-Gaussian ket.

\begin{figure}[htbp]
    \centering
    \includegraphics[scale=0.15]{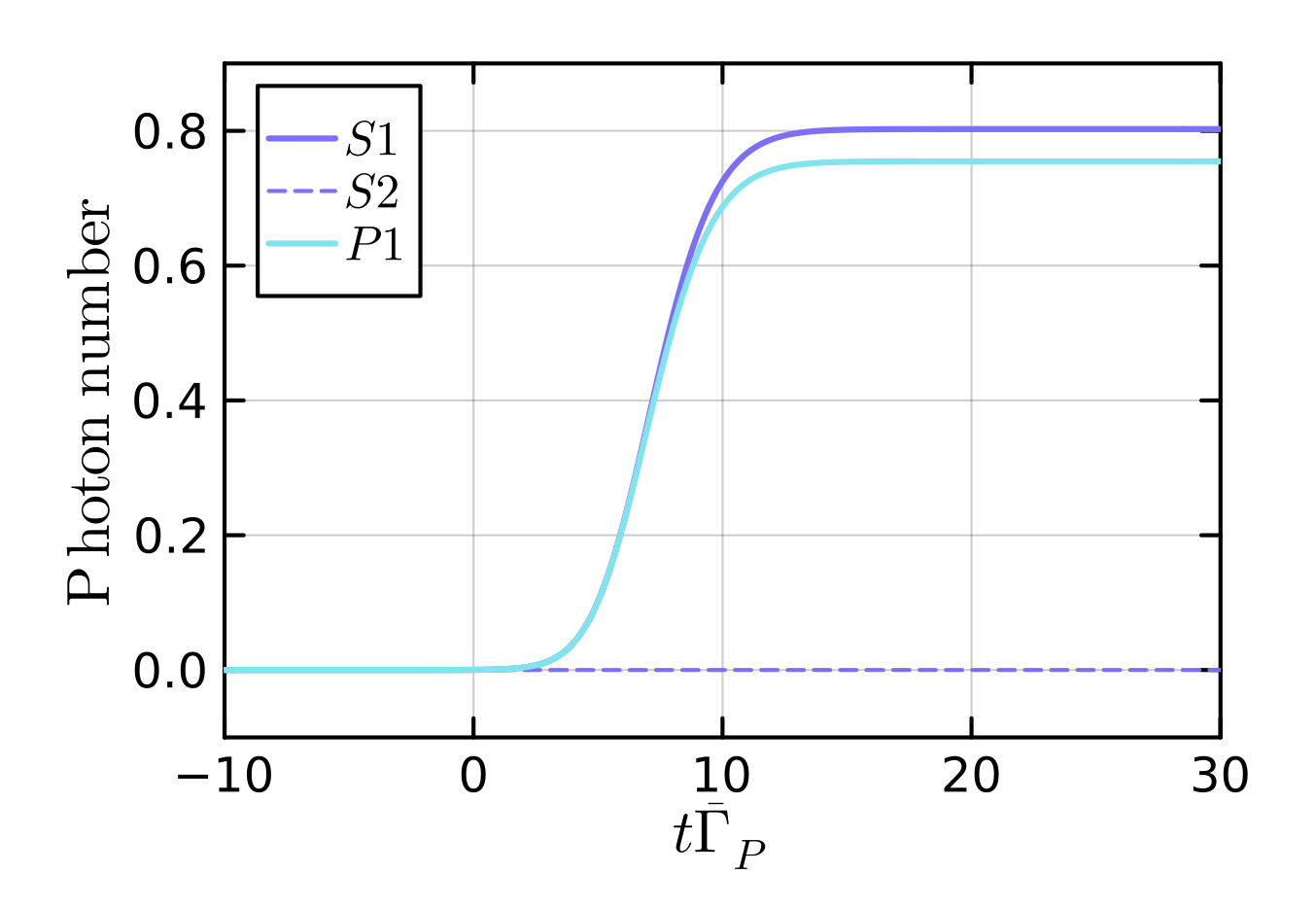}
    \caption{Non-Gaussian photon number in the first signal supermode (purple line), in the
    second signal supermode (purple dashed), and in the first pump supermode (cyan line). We scale time by the pump light decay rate, $\bar{\Gamma}_P = 2\pi \times 3.84\,{\rm GHz}$. We have taken
    $\eta_S=\eta_I = 0.95$, $Q_{{\rm int}, S} = Q_{{\rm int}, I} = 10^6$, and used a pump pulse duration of $0.5\,{\rm ns}$ and a pump energy of $0.09\,{\rm pJ}$.  }
    \label{fig:ng photon num}
\end{figure}

 To perform
 the non-Gaussian dynamical simulations we numerically solve Eq. \eqref{eq:psi tilde numerical} for
$|\tilde{\psi}(t)\rangle$ using a Fock space cutoff of $N=16$ and $\Delta t = 0.1/\overline{\Gamma}_P$. The Hamiltonian is given by Eqs. \eqref{eq:HL supermode} and \eqref{eq:Heff NG 4} for $\tilde{H}_{\rm L}(t)$ and $H_{\rm eff}(t)$ respectively, where we only keep the first pump supermode and the two leading signal and idler supermodes. Fig. \ref{fig:ng photon num} shows the non-Gaussian photon number in the first and second signal supermodes, $\langle \tilde{\psi}(t)|A^\dagger_{S1}(t)A_{S1}(t) |\tilde{\psi}(t)\rangle$ and $\langle \tilde{\psi}(t)|A^\dagger_{S2}(t)A_{S2}(t) |\tilde{\psi}(t)\rangle$, respectively. The photon number saturates at approximately $0.8$ in the first signal supermode, and at $\sim 10^{-5}$ in the second signal supermode; the idler supermode numbers are identical to the signal supermode numbers. The
first pump supermode, $\langle \tilde{\psi}(t)|B^\dagger_{P1}(t)B_{P1}(t) |\tilde{\psi}(t)\rangle$, saturates at approximately $0.75$.

\begin{figure}[htbp]
    \centering
    \includegraphics[scale=0.15]{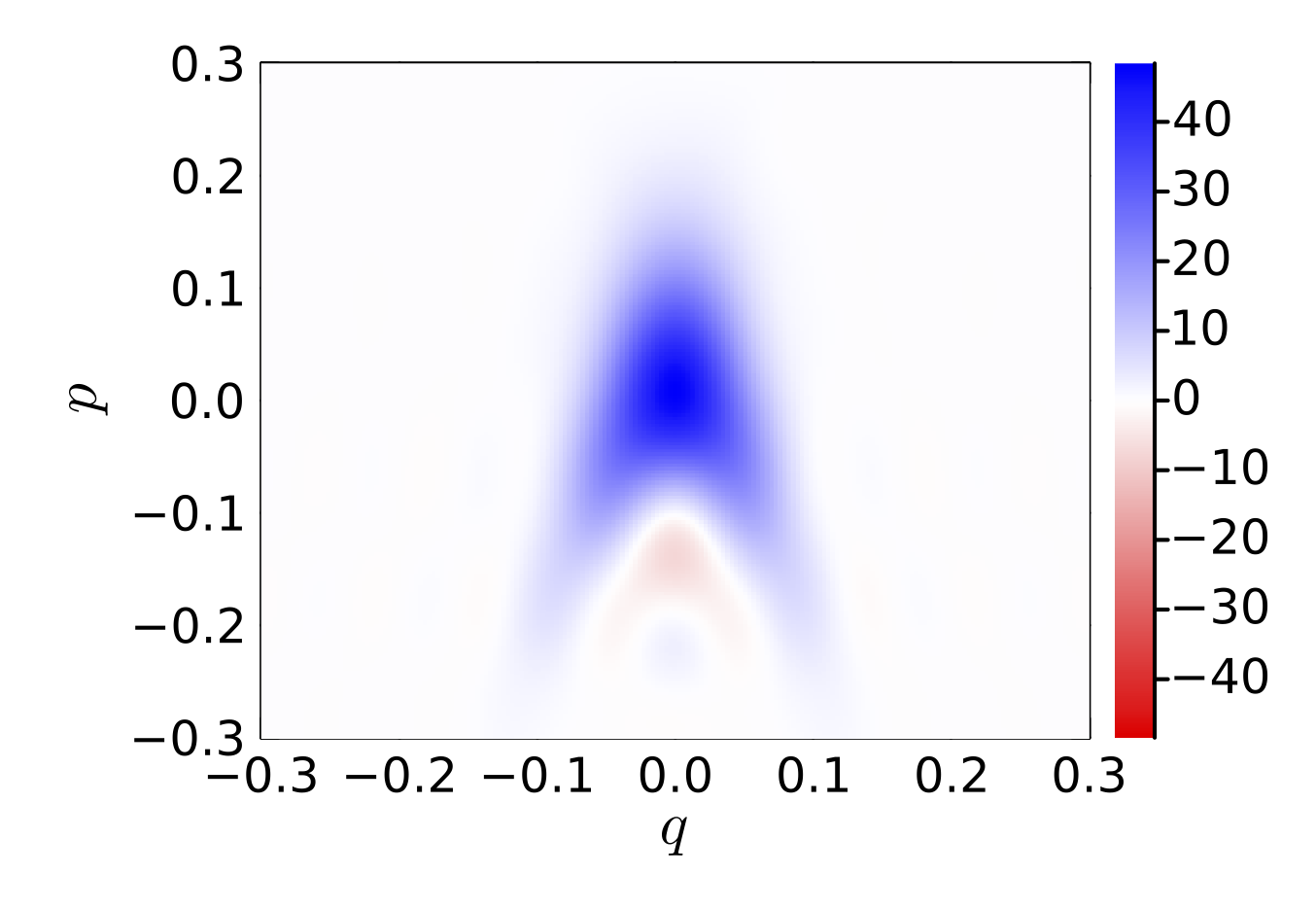}
    \caption{Wigner function for the hybrid pump, signal, and idler mode for the entire system calculated at $t_1 = 30/\bar{\Gamma}_P$, where $\bar{\Gamma}_P = 2\pi \times 3.84\,{\rm GHz}$. We used Eq. \eqref{eq:wigner},  where we put $\theta = 0.67 \pi$ and $\phi = 2.12 \pi$. We have taken
    $\eta_S=\eta_I = 0.95$, $Q_{{\rm int}, S} = Q_{{\rm int}, I} = 10^6$, a pump pulse duration of $0.5\,{\rm ns}$, and a pump energy of $0.09\,{\rm pJ}$. }
    \label{fig:wigner}
\end{figure}

These results indicate that the non-Gaussian ket contains photons in three dominant supermodes associated with the annihilation operators $A_{S1}(t)$, $A_{I1}(t)$, and $B_{P1}(t)$.  We now calculate the Wigner function for the hybrid-mode $C_{\theta, \phi}$ which is a linear combination of the dominant supermode operators \cite{Yanagimoto2022}
\begin{align}
\label{eq:hybrid mode}
    C_{\theta, \phi} = \cos(\phi) \frac{A_{S1}(t_1) + A_{I1}(t_1)}{\sqrt{2}} + \sin(\phi) {\rm e}^{i\theta}B_{P1}(t_1), 
\end{align}
where the angles $\theta$ and $\phi$ are real numbers. The Wigner function for the non-Gaussian ket is given by
\begin{align}
\label{eq:wigner}
    W^{\rm NG}_{\theta, \phi}(q,p) =  \int \frac{dxdy}{\pi^2} \langle \tilde{\psi}(t_1)|D_{\theta, \phi}(x,y)|\tilde{\psi}(t_1)\rangle{\rm e}^{2i(px - qy)},
\end{align}
where we have introduced the displacement operator $D_{\theta, \phi}(x,y)$ as
\begin{align}
    D_{\theta, \phi}(x,y) = \exp([x+iy]C^\dagger_{\theta,\phi} - {\rm H.c.}).
\end{align}

To compute Eq. \eqref{eq:wigner} for the Wigner function we calculate the expectation value $\langle \tilde{\psi}(t_1)|D_{\theta, \phi}(x,y)|\tilde{\psi}(t_1)\rangle$ for each pair of displacement variables $(x,y)$, and then do the
Fourier transform. Here $|\tilde{\psi}(t_1)\rangle$ is given by Eq. \eqref{eq:psi tilde numbers} for the ket expanded in the supermode Fock bases. 
We truncate the set of bases 
to just the three dominant basis sets, 
one each for the dominant pump, signal, and idler supermodes. All other supermodes are assumed to remain in the vacuum state. In Fig. \ref{fig:wigner} the Wigner function is shown for $\theta = 0.67 \pi$ and $\phi = 2.12 \pi$. These angles give the greatest Wigner negativity volume \cite{Kenfack2004} of approximately $0.18$, indicating that the state is non-Gaussian.

\begin{figure*}[htbp]
    \centering
\includegraphics[width=0.8\linewidth]{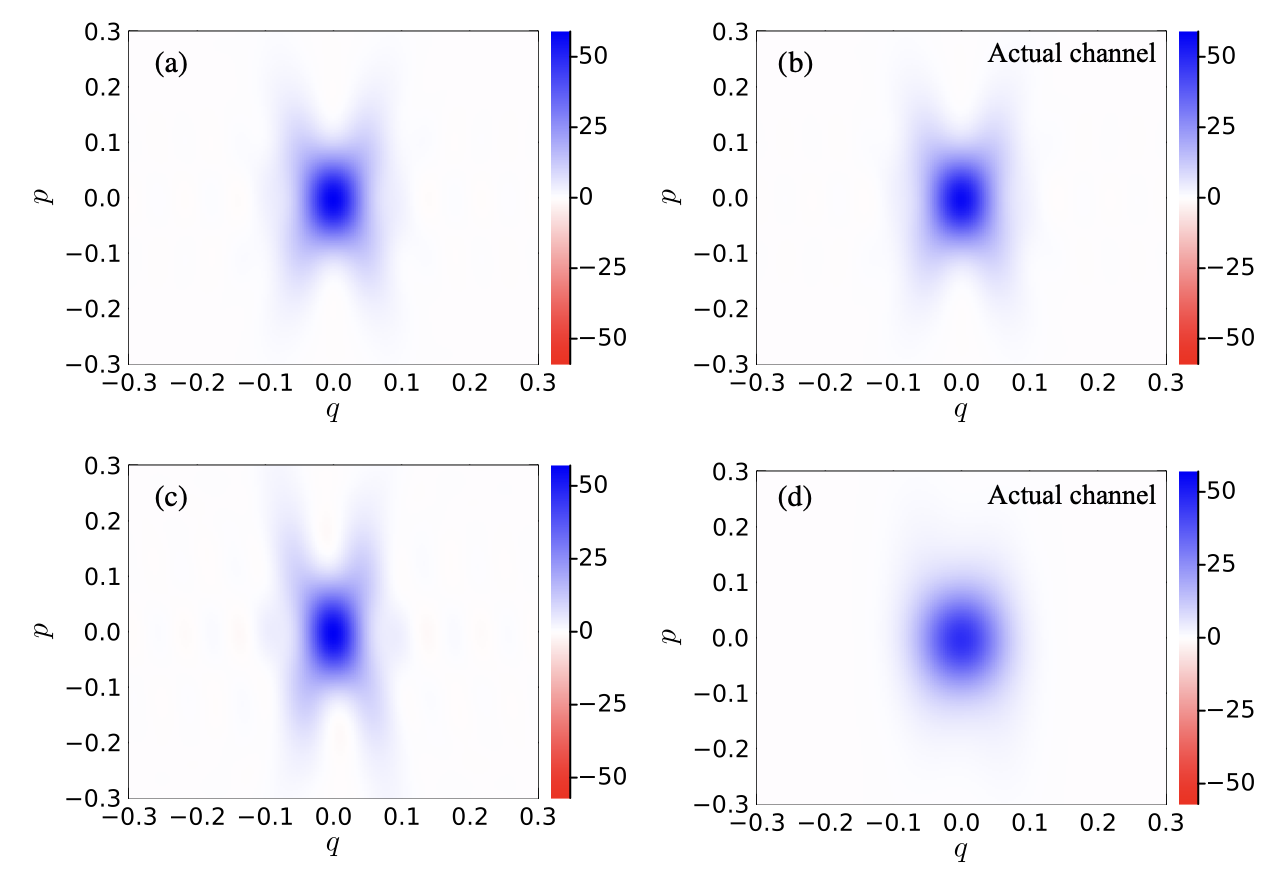}
    \caption{(a) Wigner function \eqref{eq:wigner} for the hybrid signal and idler mode $[A_{S1}(t_1) + A_{I1}(t_1)]/\sqrt{2}$ for $\eta_{S} = \eta_{I} = 0.95$ and a pump energy of $0.09\,{\rm pJ}$.  (b) Reduced Wigner function \eqref{eq:Wac} for $\eta = 0.95$ and a pump energy of $0.09\,{\rm pJ}$. (c) Wigner function \eqref{eq:wigner} for the hybrid signal and idler mode $[A_{S1}(t_1) + A_{I1}(t_1)]/\sqrt{2}$ for $\eta_{S} = \eta_{I} = 0.5$ and a pump energy of $0.01\,{\rm pJ}$.  (d) Reduced Wigner function \eqref{eq:Wac} for $\eta_{S} = \eta_{I} = 0.5$ and a pump energy of $0.01\,{\rm pJ}$. Each plot is made using 
    $Q_{{\rm int}, S} = Q_{{\rm int}, I} = 10^6$, a pump pulse duration of $0.5\,{\rm ns}$, and $t_1 = 30/\bar{\Gamma}_P$ with
    $\bar{\Gamma}_P = 2\pi \times 3.84\,{\rm GHz}$. }
    \label{fig:wigner SI}
\end{figure*}

We now consider the hybrid mode formed by a combination of the leading signal and idler supermodes for the entire system, given by $[A_{S1}(t_1) + A_{I1}(t_1)]/\sqrt{2}$. In Fig. \ref{fig:wigner SI}(a) the Wigner function for this hybrid mode for $\eta_S = \eta_I = 0.95$ and a pump energy of $0.09\,{\rm pJ}$ is shown, which displays non-Gaussian features. To investigate the effect of loss we compute the reduced Wigner function for the actual channel only. In Appendix \ref{sec:actual channel wigner} we show that the reduced Wigner function for $\eta_S = \eta_I \equiv \eta$ is well approximated by
 \begin{align}
 \label{eq:Wac}
    W^{\rm NG}_{\rm ac}(q,p) &\approx  \int \frac{dxdy}{\pi^2} \nonumber
    \\
    &\times \langle \tilde{\psi}|\exp(\sqrt{\eta}\,\frac{x+iy}{\sqrt{2}}\left[A^\dagger_{S1}+ A^\dagger_{I1}\right]- {\rm H.c.})|\tilde{\psi}\rangle\nonumber
    \\
    &\times \exp(-(1-\eta)\,\frac{x^2+y^2}{2}){\rm e}^{2i(px - qy)},
\end{align}
where we dropped the time-dependence on the ket for convenience.
The integrand in Eq. \eqref{eq:Wac} contains a vacuum noise term that depends on the loss as $1-\eta$. When there is loss ($\eta<1$) the vacuum noise term can dominate the Wigner function, causing it to resemble that of the vacuum state. 

In Fig. \ref{fig:wigner SI}(b) the reduced Wigner function is shown for $\eta = 0.95$ and a pump energy of $0.09\,{\rm pJ}$. Here
the non-Gaussian features persist in the reduced Wigner function. Compare this with the scenario shown in Fig. \ref{fig:wigner SI}(c) and Fig. \ref{fig:wigner SI}(d). In Fig. \ref{fig:wigner SI}(c) we plot the Wigner function for the signal and idler hybrid mode $[A_{S1}(t_1) + A_{I1}(t_1)]/\sqrt{2}$ if we assume
critical coupling ($\eta_S = \eta_I = 0.5$) and a pump energy of $0.01\,{\rm pJ}$, and in
Fig. \ref{fig:wigner SI}(d) we plot the reduced Wigner function that results.
Here tracing out the phantom channel erases the non-Gaussian features, leading to a reduced Wigner function for the actual channel that resembles the Wigner function of the vacuum state.

\subsection{Stretching the full ket in the actual channel}
In the previous subsection we calculated the Wigner function for the non-Gaussian ket. In practice one does not have access to the non-Gaussian ket itself. To model a realistic measurement we calculate the expectation value of observables involving the actual channel operators with respect to the full ket. We show that its non-Gaussian features are obscured by the large squeezing amplitude, but they can be revealed by applying a unitary
to the state
in the actual channel. 

For microring resonator systems pumped with a coherent state, strong squeezing is required to achieve sufficient pump depletion to observe non-Gaussianity. If one calculates the Wigner function of the hybrid mode $[A_{S1}(t_1) + A_{I1}(t_1)]/\sqrt{2}$ using the full ket, the phase-space portrait in Fig. \ref{fig:wigner SI}(b) for the reduced Wigner function in the actual channel is squeezed along the $q$-axis by a factor $\exp(-r_1) \approx 0.025$, where the squeezing parameter is $r_1 = 3.7$. This extreme squeezing obscures the non-Gaussian features.

In principle, the Gaussian features in the Wigner function of the full ket can be removed by applying the inverse Gaussian transformations to the light after the ring. Prior to any measurement, one could imagine applying a Gaussian unitary that would stretch the Wigner function
along the  $q$-axis by a factor of $\exp(r_1)$, cancelling
the initial squeezing. An appropriate rotation and displacement of the ket could
also be applied 
if needed. These operations would be
equivalent to applying the Hermitian conjugate of the Gaussian unitary $U^\dagger$ to the full ket after the ring, leading to a final full ket equal to the non-Gaussian ket,
$U^\dagger \ket{\psi} = |\tilde{\psi}\rangle$. The Wigner function of the final full ket would then be equal to that of the non-Gaussian ket.

\begin{figure}[htbp]
    \centering
    \includegraphics[width=.7\linewidth]{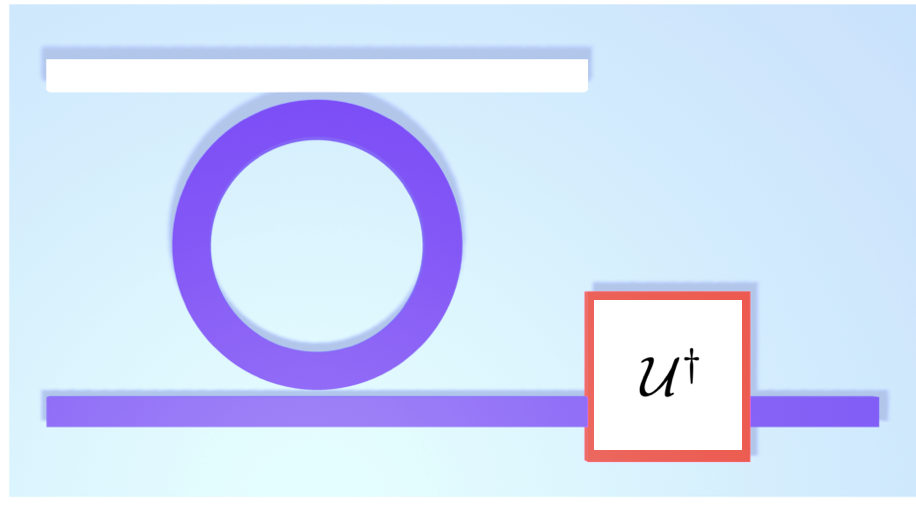}
    \caption{Schematic of the system used to isolate the non-Gaussian features of the full ket. A Gaussian unitary $\mathcal{U^\dagger}$ is applied only to the light in the actual output channel.}
    \label{fig:undo squeezing}
\end{figure}

In practice it is impossible to implement $U^\dagger$ exactly, since it requires access to scattered photons in the phantom channel. One can only implement a unitary $\mathcal{U}^\dagger$ that acts on the light in the actual channel. In Fig. \ref{fig:undo squeezing} is a schematic of the system that implements $\mathcal{U}^\dagger$ on the light exiting the ring in the actual channel. The final ket can be written as $\ket{\psi'} = \mathcal{U}^\dagger U|\tilde{\psi}\rangle$. The squeezing parameters of $\mathcal{U}^\dagger$ should be chosen so that the combined unitary $\mathcal{U}^\dagger U$ minimizes squeezing of the light in the actual channel. 
We employ a strategy in which the Gaussian unitary is initially calculated under the artificial condition of no scattering loss, $U^\dagger_{\rm no\,loss}$. In this case, we can set the unitary in the actual channel to 
$\mathcal{U}^\dagger = U^\dagger_{\rm no\,loss}$, ensuring that the Gaussian operations in the ket are exactly inverted. When scattering loss is introduced we expect that a reasonable choice for $\mathcal{U}^\dagger$ would be $\mathcal{U}^\dagger \approx U^\dagger_{\rm no\, loss}$ provided the loss is not too large. We investigated this strategy below.

\begin{figure*}[htbp]
    \centering
    \includegraphics[width=0.8\linewidth]{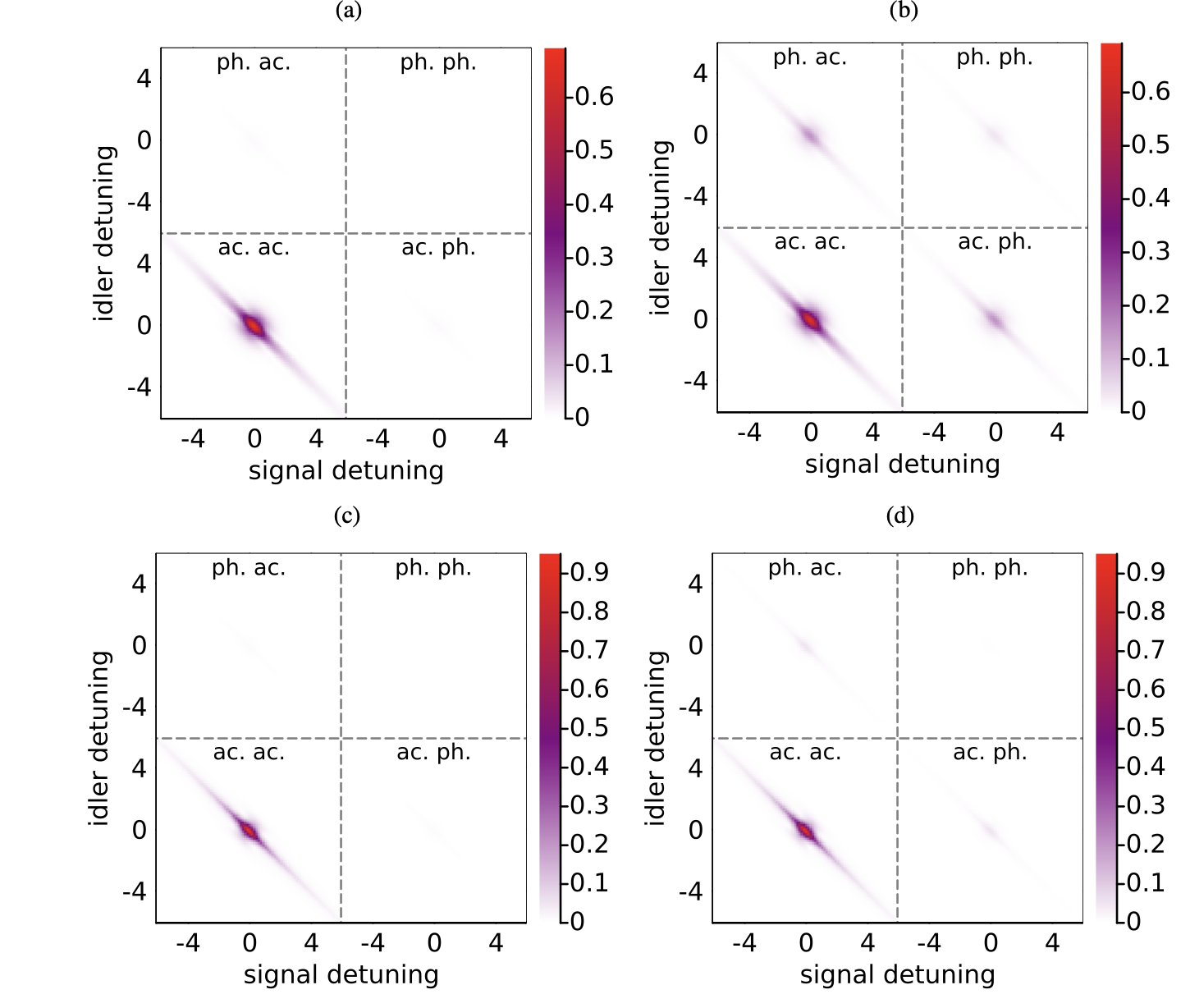}
    \caption{Absolute values of the squeezing matrix elements $|[\bm J(t_1)]{\mu_1\mu_2}|$ (see Eq. \eqref{eq:S}) at a late time $t_1$ after the pump has left the ring, where $\mu_1 = ({\rm ac.}, k_1)$ or $({\rm ph.}, k_1)$ denotes photons exiting the actual or phantom channel, respectively. Signal and idler detunings are defined as $(k_1-K_S)v_S/\bar{\Gamma}_S$ and $(k_2-K_I)v_I/\bar{\Gamma}_I$, with $K_J$ the resonance center, $v_J$ the group velocity, and $2\bar{\Gamma}_J$ the linewidth ($J=S,I$). (a) No signal/idler loss: pump energy $0.09\,{\rm pJ}$, pulse duration $0.5\,{\rm ns}$, $\eta_P=0.5$, $Q_{{\rm int},P}=10^5$. (b) With signal/idler loss: $\eta_S=\eta_I=0.95$, $Q_{{\rm int},S}=Q_{{\rm int},I}=10^6$, pump energy $0.09\,{\rm pJ}$, pulse duration $0.5\,{\rm ns}$, $\eta_P=0.5$, $Q_{{\rm int},P}=10^5$. (c) No signal/idler loss: pump energy $0.027\,{\rm pJ}$, pulse duration $0.3\,{\rm ns}$, $\eta_P=0.5$, $Q_{{\rm int},P}=10^6$. (d) With signal/idler loss: $\eta_S=\eta_I=0.997$, $Q_{{\rm int},S}=Q_{{\rm int},I}=15\times10^6$, pump energy $0.027\,{\rm pJ}$, pulse duration $0.3\,{\rm ns}$, $\eta_P=0.5$, $Q_{{\rm int},P}=10^6$.}
    \label{fig:J actual channel}
\end{figure*}

The full Gaussian unitary $U$ contains a squeezing matrix $\bm J(t)$ (see Appendix \ref{sec:U(t)}). In Fig. \ref{fig:J actual channel}(a) we plot the absolute value of the elements of the squeezing matrix $|[\bm J(t_1)]_{\mu_1\mu_2}|$  for the artificial case where there is no signal and idler loss. In this lossless case the components of $|[\bm J(t_1)]_{\mu_1\mu_2}|$ involving the phantom channel are zero. Now we include signal and idler loss in Fig. \ref{fig:J actual channel}(b), where $|[\bm J(t_1)]_{\mu_1\mu_2}|$ is shown for $\eta_S = \eta_I = 0.95$ and $Q_{{\rm int},S} = Q_{{\rm int},I} = 10^6$. The loss causes the phantom channel components of $|[\bm J(t_1)]_{\mu_1\mu_2}|$ to be non-zero. We can see that the non-zero phantom channel components prevent the squeezing matrix from being identical to the lossless squeezing matrix in Fig. \ref{fig:J actual channel}(a).  

\begin{figure*}[htbp]
    \centering
    \includegraphics[scale=.56]{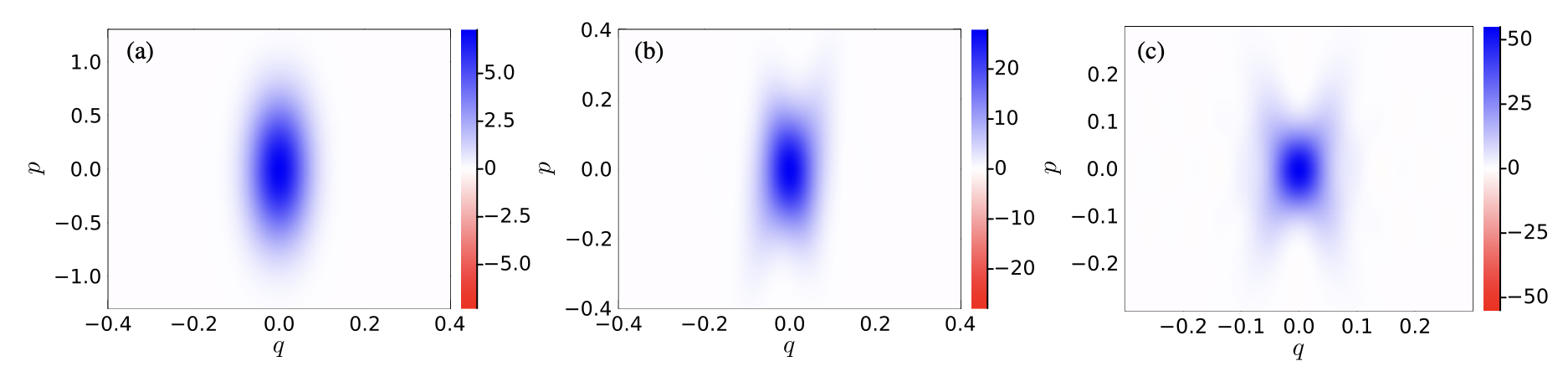}
    \caption{Wigner functions for the full ket in the actual channel after a Gaussian unitary $\mathcal{U}^\dagger$ is applied (see schematic in Fig. \ref{fig:undo squeezing}). (a) High-loss scenario using the squeezing matrix in Fig. \ref{fig:J actual channel}(a) to construct $\mathcal{U}^\dagger$: pump energy $0.09\,{\rm pJ}$, pulse duration $0.5\,{\rm ns}$, $\eta_P = 0.5$, $Q_{{\rm int}, P} = 10^5$, $\eta_S =\eta_I = 0.95$, $Q_{{\rm int}, S}= Q_{{\rm int}, I} = 10^6$, and $r_1 = 3.7$. (b) Low-loss scenario using the squeezing matrix in Fig. \ref{fig:J actual channel}(c) to construct $\mathcal{U}$: pump energy $0.027\,{\rm pJ}$, pulse duration $0.3\,{\rm ns}$, $\eta_P = 0.5$, $Q_{{\rm int}, P} = 10^6$, $\eta_S =\eta_I = 0.997$, $Q_{{\rm int}, S}= Q_{{\rm int}, I} = 15\times10^6$, and $r_1 = 3.6$. Artificial scenario with arbitrarily low signal/idler loss: (c) pump energy $0.09\,{\rm pJ}$, pulse duration $0.5\,{\rm ns}$, $\eta_P = 0.5$, $Q_{{\rm int}, P} = 10^5$, $\eta_S =\eta_I = 0.9999$, $Q_{{\rm int}, S}= Q_{{\rm int}, I} = 500\times10^6$, and $r_1 = 3.7$.}
    \label{fig:full ket wigner}
\end{figure*}

We are now in a position to implement the strategy outlined above. We construct $\mathcal{U}^\dagger$ using the (ac., ac.) component of the squeezing matrix given in Fig. \ref{fig:J actual channel}(a) and calculate the Wigner function for $\ket{\psi'} = \mathcal{U}^\dagger U|\tilde{\psi}\rangle$. The details are in Appendix \ref{sec:undo squeezing wigner}, here we just present the results. In Fig. \ref{fig:full ket wigner}(a) we show the resulting Wigner function for $\ket{\psi'}$. The state resembles a squeezed vacuum with an effective squeezing parameter of approximately $1$. So applying $\mathcal{U}^\dagger$ reduces the squeezing parameter from $3.7$ to about $1$. Much of the squeezing is removed but the phantom channel components of $\bm J(t)$ prevent sufficient stretching along the $q$-axis to reveal the non-Gaussian features in the Wigner function.

With these parameters an effective stretching to reveal the non-Gaussian features cannot be done. To move to a regime where it is possible, we consider an increase in the signal and idler escape efficiency ($\eta_{S} = \eta_I = 0.997$) and the intrinsic quality factors for the signal, idler, and pump ($Q_{{\rm int}, S} =  Q_{{\rm int}, I} = 15\times 10^{\rm 6}$, and $Q_{{\rm int}, P} = 10^6$). To maintain a similar squeezing level we choose a pump energy and duration of $0.027\,{\rm pJ}$ and $0.3 \,{\rm ns}$, respectively. 

Using these new parameters, we proceed in a similar way as above and first consider $|[\bm J(t_1)]_{\mu_1\mu_2}|$ in Fig. \ref{fig:J actual channel}(c) for the artificial case of no signal and idler loss. Then signal and idler losses are introduced in Fig. \ref{fig:J actual channel}(d), where now the magnitude of the phantom channel elements are visibly smaller than those in Fig. \ref{fig:J actual channel}(b). We construct $\mathcal{U}^\dagger$ using the (ac., ac.) component of $\bm J(t)$ given in Fig. \ref{fig:J actual channel}(c) and calculate the Wigner function for $\ket{\psi'} = \mathcal{U}^\dagger U|\tilde{\psi}\rangle$. In Fig. \ref{fig:full ket wigner}(b) we show the resulting Wigner function, where now a sufficient amount of squeezing has been removed to reveal the non-Gaussian features. For comparison, in Fig. \ref{fig:full ket wigner}(c) we show that the squeezing could be entirely removed under the artificial conditions where the escape efficiencies approach unity and the intrinsic quality factors approach infinity.

\section{Conclusion}
\label{sec:conclusion}
We have developed a framework for modeling non-Gaussian state generation via SPDC
that incorporates scattering loss with a phantom channel. 
The full ket for the system is written as a Gaussian unitary acting on a residual non-Gaussian ket, separating the evolution into Gaussian and non-Gaussian parts. In this framework we are able to use the Gaussian solution to find supermodes that enable the efficient simulation of the non-Gaussian ket. 

Using this approach, in calculations using realistic parameters for InGaP microrings we found
a Wigner negativity volume of approximately $0.18$ for the residual non-Gaussian ket. While
the non-Gaussian features are largely obscured in the Wigner function of the full ket by the squeezing, they could
be revealed by applying an appropriate inverse Gaussian unitary that undid the squeezing by ``stretching." 
Although the intrinsic quality factors needed for this approach are
roughly an order of magnitude higher than those currently being reported in the
InGaP literature, with future design and fabrication improvements this should be a viable approach.

 In the InGaP microrings considered, the onset of non-Gaussianity required high escape efficiencies and pump energies on the order of $0.1\,{\rm pJ}$,  yielding a Wigner function that was squeezed along one axis
 by a factor of $\sim0.025$. In contrast, PPLN waveguide models require less pump energy, typically achieving non-Gaussianity with a comparatively smaller squeezing factor (e.g., $\sim 0.1$) \cite{Yanagimoto2022}. Although waveguide models achieve non-Gaussianity with comparatively less squeezing, 
 both microring and waveguide systems share the challenge that the
 ability to reveal non-Gaussian features depends on suppressing the dominant Gaussian features in the Wigner function. In addition, we showed that the suppression of Gaussian features critically depends on the accurate inclusion of scattering loss in the model. Ultimately, the advantage of microring resonators are their compactness, and with effective handling of scattering loss, they open a path for the deterministic generation of non-Gaussian states in a compact and scalable architecture.

\section*{Acknowledgements}
This work is supported by Mitacs through the Mitacs accelerate program Grant No. IT15614.
The authors acknowledge Natural  Sciences  and  Engineering  Research Council of Canada (NSERC) for funding. S.E.F. and J.E.S. acknowledge the Natural Sciences and Engineering Research Council of Canada European Union’s Horizon Europe Research and Innovation Programme (101070700, project MIRAQLS) for financial support. S.E.F. acknowledges support from a Walter C. Sumner Memorial Fellowship.

\bibliography{References-minimum} 

\appendix
\section{Gaussian unitary}
\label{sec:U(t)}

In this section we define the Gaussian unitary that identifies the Gaussian dynamics of the full ket, and we relate its squeezing and rotation parameters to the solution of Eqs. \eqref{eq:Wdot} and \eqref{eq:Vdot}.

The Gaussian unitary is written as
\begin{align}
\label{eq:U}
    U(t) = {\rm e}^{i\gamma(t)}D[\bm x(t)]S[\bm J(t)]R[\bm \phi(t)],
\end{align}
where the phase $\gamma(t)$ is a real function of time, and the unitary displacement, squeezing, and rotation operators are given by
\begin{align}
\label{eq:D}
    D[\bm x(t)]&=\exp([\bm x(t)]_{\mu_3} b^\dagger_{P\mu_3} - {\rm H.c.} ),
    \\
    \label{eq:S}
    S[\bm J(t)]&=\exp([\bm J(t)]_{\mu_1 \mu_2} a^\dagger_{S\mu_1}a^\dagger_{I\mu_2} - {\rm H.c.} ),
    \\
    \label{eq:R}
    R[\bm \phi(t)]&=\exp(i[\bm \phi_S(t)]_{\mu_1 \mu_2} a^\dagger_{S\mu_1}a_{S\mu_2})\nonumber
    \\
    &\times \exp(i[\bm \phi_I(t)]_{\mu_1 \mu_2} a^\dagger_{I\mu_1}a_{I\mu_2}),
\end{align}
where $\bm x(t)$ is a vector of displacements for the pump, $\bm J(t)$ is the non-degenerate squeezing matrix, and $\bm \phi_S(t)$ and $\bm \phi_I(t)$ are the Hermitian rotation matrices for the signal and idler, respectively. 

Putting Eq. \eqref{eq:U} for $U(t)$ into the RHS of Eq. \eqref{eq:idler trans} and the RHS of Eq. \eqref{eq:signal trans} for $U^\dagger(t) a_{I\mu_1} U(t)$ and $U^\dagger(t) a_{S\mu_1} U(t)$, respectively, and equating coefficients in front of similar operators on both sides we obtain
\begin{align}
\label{eq:VSS decomp}
    \bm V_{SS}(t) &= \cosh[\bm u(t)]\exp[i\bm \phi_S(t)],
    \\
\label{eq:VII decomp}
     \bm V_{II}(t) &= \cosh[\bm u(t)]\exp[i\bm \phi_I(t)],
     \\
\label{eq:WIS decomp}
      \bm W_{IS}(t) &= \sinh[\bm u(t)]\exp[i\bm \alpha(t)]\exp[-i\bm \phi^*_S(t)],
          \\
\label{eq:WSI decomp}
      \bm W_{SI}(t) &= \sinh[\bm u(t)]\exp[i\bm \alpha(t)]\exp[-i\bm \phi^*_I(t)],
\end{align}
where the Hermitian matrices $\bm u(t)$ and $\bm \alpha(t)$ are from the polar decomposition of the squeezing matrix
\begin{align}
\label{eq:J}
    \bm J(t) = \bm u(t) \exp[i\bm \alpha(t)].
\end{align}
Eqs. \eqref{eq:VSS decomp} - \eqref{eq:WSI decomp} give the solution of Eqs. \eqref{eq:Wdot} and \eqref{eq:Vdot} in terms of the squeezing matrix and rotation matrices. 

We now show how to extract the squeezing matrix $\bm J(t)$ and rotation matrices $\bm \phi_S(t)$ and $\bm \phi_I(t)$ from Eqs. \eqref{eq:VSS decomp} - \eqref{eq:WSI decomp}.  We closely follow our previous strategy \cite{Vendromin2024}.  Eqs. \eqref{eq:VSS decomp} and \eqref{eq:VII decomp} for $\bm V_{SS}(t)$ and $\bm V_{II}(t)$ are in the form of polar decompositions already, where $\cosh[\bm u(t)]$ is a Hermitian positive semidefinite (PSD) matrix and $\exp[i\bm \phi_S(t)]$ and $\exp[i\bm \phi_I(t)]$ are unitary matrices.  

Having uniquely identified the matrices $\cosh[\bm u(t)]$, $\exp[i\bm \phi_S(t)]$ and $\exp[i\bm \phi_I(t)]$ from the polar decompositions, the SVD of $\cosh[\bm u(t)]$ can be written as
\begin{align}
\label{eq:coshu svd}
    \cosh[\bm u(t)] = \bm F(t) \cosh[\bm \sigma(t)] \bm F^\dagger(t),
\end{align}
where $\bm \sigma(t) ={\rm diag}[\sigma_1(t), \ldots]$ is a diagonal matrix, and
\begin{align}
\label{eq:u svd}
    \bm u(t) = \bm F(t) \bm \sigma(t) \bm F^\dagger(t),
\end{align}
with the unitary matrix $\bm F(t)$ the same in both decompositions. Using  Eq. \eqref{eq:u svd} for $\bm u(t)$ we can write
\begin{align}
   \label{eq:sinhu svd}
    \sinh[\bm u(t)] = \bm F(t) \sinh[\bm \sigma(t)] \bm F^\dagger(t).
\end{align}
Putting Eq. \eqref{eq:sinhu svd} for $\sinh[\bm u(t)]$ into Eq. \eqref{eq:WIS decomp} for $\bm W_{IS}(t)$ 
we obtain
\begin{align}
\label{eq:WIS decomp 2}
    \bm W_{IS}(t) &= \bm F(t) \sinh[\bm \sigma(t)] \bm F^\dagger(t)\exp[i\bm \alpha(t)]\exp[-i\bm \phi^*_S(t)].
\end{align}
Introducing the matrix $\bm K(t)$ as
\begin{align}
    \bm K(t) = \bm F(t) \sinh[\bm \sigma(t)] \bm \sigma^{-1}(t) \bm F^\dagger(t),
\end{align}
and its inverse 
\begin{align}
    \bm K^{-1}(t) = \bm F(t)  \bm \sigma(t)\{\sinh[\bm \sigma(t)]\}^{-1} \bm F^\dagger(t),
\end{align}
we can multiply Eq. \eqref{eq:WIS decomp 2} from the left by $\bm K^{-1}(t)$ to obtain
\begin{align}
\label{eq:WIS decomp 3}
   \bm K^{-1}(t) \bm W_{IS}(t) &= \bm J(t)\exp[-i\bm \phi^*_S(t)].
\end{align}
So the squeezing matrix $\bm J(t)$ is given by
\begin{align}
\label{eq:J solved}
    \bm J(t) = \bm K^{-1}(t) \bm W_{IS}(t) \exp[i\bm \phi^*_S(t)].
\end{align}
All quantities on the right-hand side of Eq. \eqref{eq:J solved} are known, so we can determine $\bm J(t)$. 

We stress that the vector of displacements for the pump $\bm x(t)$ is obtained directly from the solution of the coupled differential equations \eqref{eq:xdot}, \eqref{eq:Wdot}, and \eqref{eq:Vdot}.

\section{Removing Gaussian terms in $H_{\rm eff}(t)$}
\label{sec:heff}

In this section we simplify Eq. \eqref{eq:Heff} for $H_{\rm eff}(t)$ by removing all Gaussian terms. 

The term that needs to be simplified in Eq. \eqref{eq:Heff} is the Gaussian transformation of the nonlinear Hamiltonian, $U^\dagger(t) H_{\rm NL}(t) U(t)$. We begin by evaluating $U^\dagger(t) H_{\rm NL}(t) U(t)$, where $H_{\rm NL}(t)$ is given by Eq. \eqref{eq:HNL} in the interaction picture
\begin{align}
\label{eq:UHNLU}
    U^\dagger(t) H_{\rm NL}(t) U(t)&= \hbar[\Lambda(t)]_{\mu_1 \mu_2 \mu_3} U^\dagger(t) a^\dagger_{S\mu_1} a^\dagger_{I\mu_2} U(t) \nonumber
    \\
    &\times U^\dagger(t)b_{P\mu_3} U(t) \nonumber
    \\
    &+ {\rm H.c.},
\end{align}
where the explicit notation for the sum over indices is removed for convenience throughout this section. Using Eqs.  \eqref{eq:signal trans}, and \eqref{eq:idler trans} for $U^\dagger(t) a_{S\mu_1}U(t)$, and  $U^\dagger(t) a_{I\mu_1}U(t)$ we can write
\begin{align}
\label{eq:UaaU}
U^\dagger(t) a^\dagger_{S\mu_1} a^\dagger_{I\mu_2} U(t)&= [\bm V^*_{SS}(t)]_{\mu_1 \mu'_1} [\bm V^\dagger_{II}(t)]_{\mu'_2 \mu_2} a^\dagger_{S\mu'_1}a^\dagger_{I\mu'_2} \nonumber
\\
&+[\bm V^*_{SS}(t)]_{\mu_1 \mu'_1} [\bm W^\dagger_{IS}(t)]_{\mu'_2 \mu_2} a^\dagger_{S\mu'_1}a_{S\mu'_2} \nonumber
\\
&+[\bm W^*_{SI}(t)]_{\mu_1 \mu'_1} [\bm V^\dagger_{II}(t)]_{\mu'_2 \mu_2} a^\dagger_{I\mu'_2}a_{I\mu'_1} \nonumber
\\
&+[\bm W^*_{SI}(t)]_{\mu_1 \mu'_1} [\bm W^\dagger_{IS}(t)]_{\mu'_2 \mu_2} a_{I\mu'_1}a_{S\mu'_2} \nonumber
\\
&+[\bm W^*_{SI}(t)\bm V^\dagger_{II}(t)]_{\mu_1 \mu_2}.
\end{align}
Putting Eq. \eqref{eq:pump trans} for $U^\dagger(t)b_{P\mu_3} U(t)$ and Eq. \eqref{eq:UaaU} for $U^\dagger(t) a^\dagger_{S\mu_1} a^\dagger_{I\mu_2} U(t)$ into Eq. \eqref{eq:UHNLU} above, we find it can be written as
\begin{align}
\label{eq:UHNLU 2}
    U^\dagger(t) H_{\rm NL}(t) U(t)&=U^\dagger(t) H_{\rm G}(t) U(t) + H_{\rm NG}(t),
\end{align}
where we have introduced the Gaussian Hamiltonian
\begin{align}
    \label{eq:HG}
    H_{\rm G}(t)&=\hbar [\bm Z(t)]_{\mu_1 \mu_2 }a^\dagger_{S\mu_1}a^\dagger_{I\mu_2}  \nonumber
\\
&+ \hbar [\bm \Delta^*(t)]_{ \mu_3}  \{ b_{P\mu_3} -[\bm x(t)]_{\mu_3}\} \nonumber
\\
&+ {\rm H.c.},
\end{align}
where $\bm Z(t)$ and $\bm \Delta(t)$ are given by Eqs. \eqref{eq:Z} and \eqref{eq:Delta} respectively, and the non-Gaussian Hamiltonian $H_{\rm NG}(t)$ is given by Eq. \eqref{eq:Heff NG}. 

We now put Eq. \eqref{eq:UHNLU 2} for the simplified $U^\dagger(t) H_{\rm NL}(t) U(t)$ into Eq. \eqref{eq:Heff} for $H_{\rm eff}(t)$ 
\begin{align}
\label{eq:Heff trans workout}
    H_{\rm eff}(t)&=  U^\dagger(t) H_{\rm G}(t) U(t) + H_{\rm NG}(t)- i\hbar U^\dagger(t) \frac{dU(t)}{dt} ,
\end{align}
To cancel all the Gaussian terms in Eq. \eqref{eq:Heff trans workout} clearly we just need to enforce that $U(t)$ is a solution to the Schr\"odinger equation
\begin{align}
\label{eq:Gaussian SE}
    i\hbar \frac{dU(t)}{dt} = H_{\rm G}(t) U(t).
\end{align}
Enforcing this condition leads to Eq. \eqref{eq:Gaussian SE} makes Eq. \eqref{eq:Heff trans workout} for  $H_{\rm eff}(t)$ containing only non-Gaussian terms
\begin{align}
\label{eq:Heff purely ng}
    H_{\rm eff}(t) = H_{\rm NG}(t).
\end{align}

Having simplified $H_{\rm eff}(t)$ we now derive Eqs. \eqref{eq:xdot}, \eqref{eq:Vdot}, and \eqref{eq:Wdot} for the coupled differential equations for the Gaussian parameters of $U(t)$. Putting Eq. \eqref{eq:U} for the displaced, squeezed, rotated form of $U(t)$ into Eq. \eqref{eq:Gaussian SE} and multiplying from the left by $U^\dagger(t)$
we obtain
\begin{align}
\label{eq:dVdt}
    i\hbar \mathcal{V}^\dagger(t)\frac{d\mathcal{V}(t)}{dt} = \mathcal{V}^\dagger(t)H_{\rm G}(t) \mathcal{V}(t)+\hbar\frac{d\gamma(t)}{dt},
\end{align}
where we have introduced the unitary operator $\mathcal{V}(t)$ as
\begin{align}
\label{eq:V(t)}
    \mathcal{V}(t) = D[\bm x(t)]S[\bm J(t)]R[\bm \phi(t)].
\end{align}
A differential equation for the global phase $\gamma(t)$ can be derived by requiring that $d\gamma(t)/dt$ cancels all terms proportional to the identity in Eq. \eqref{eq:dVdt}. This approach was originally introduced by Ma and Rhodes \cite{marhodes1990} and has recently been extended by Quesada to a more practical and scalable method for computing the phase \cite{Quesada2025}. The global phase $\gamma(t)$ does not affect any of our results, so we put $\gamma(t) = 0$  and neglect terms proportional to the identity in Eq. \eqref{eq:dVdt}.  Then Eq. \eqref{eq:dVdt} can be written as
\begin{align}
\label{eq:dVdt 2}
    i\hbar\frac{d\mathcal{V}(t)}{dt} = H_{\rm G}(t) \mathcal{V}(t).
\end{align}

From Ma and Rhodes \cite{marhodes1990} we know that since $H_{\rm G}(t)$ is at most quadratic in operators, the unique solution (neglecting the global phase) to Eq. \eqref{eq:dVdt 2} is given by a squeezed, displaced, rotated unitary given by Eq. \eqref{eq:V(t)} for $\mathcal{V}(t)$. The task is then to obtain equations for the Gaussian parameters $\bm x(t)$, $\bm J(t)$, and $\bm \phi(t)$. To obtain these equations we borrow a strategy from previous work \cite{Quesada2022, Vendromin2024}. Consider the vector of time-dependent operators
\begin{align}
\label{eq:a(t)}
    \begin{bmatrix}
        \bm a_S(t)
        \\
        \bm a_I(t)
    \end{bmatrix}
    &\equiv 
     \begin{bmatrix}
        U^\dagger(t) \bm a_S U(t)
        \\
        U^\dagger(t) \bm a_I U(t)
    \end{bmatrix} \nonumber
=
    \begin{bmatrix}
        \bm V_{SS}(t) & 0
        \\
        0 & \bm V_{II}(t)
    \end{bmatrix}
\begin{bmatrix}
        \bm a_S
        \\
        \bm a_I
    \end{bmatrix}
    \\
    &+
    \begin{bmatrix}
        0&\bm W_{SI}(t)
        \\
        \bm W_{IS}(t) & 0
    \end{bmatrix}
\begin{bmatrix}
        \bm a^\dagger_S
        \\
        \bm a^\dagger_I
    \end{bmatrix},
\end{align}
where for convenience we have introduced the column vector of operators $\bm a_J = [ a_{J1}, a_{J2}, \ldots]^{\rm T}$ for $J=S, I$ and T is the transpose, and we have used Eqs. \eqref{eq:signal trans} and \eqref{eq:idler trans} for $U^\dagger(t) \bm a_S U(t)$ and $U^\dagger(t) \bm a_I U(t)$.  Taking the time derivative of Eq. \eqref{eq:a(t)} we obtain
\begin{align}
\label{eq:daSdt}
\frac{d\bm a_S(t)}{dt}&= \frac{d\bm V_{SS}(t)}{dt} \bm a_S +\frac{d\bm W_{SI}(t)}{dt} \bm a^\dagger_I,
        \\
\label{eq:daIdt}
\frac{d\bm a_I(t)}{dt}&=\frac{d\bm V_{II}(t)}{dt} \bm a_I +\frac{d\bm W_{IS}(t)}{dt} \bm a^\dagger_S.
\end{align}
We can form an equivalent expression for the derivative of the operators $\bm a_S$ and $\bm a_I$ with the Heisenberg picture using Eq. \eqref{eq:HG} for the Gaussian Hamiltonian. In the Heisenberg picture the derivatives of the operators are
\begin{align}
\label{eq:heisenberg eq}
    \begin{bmatrix}
        \frac{d\bm a_S(t)}{dt}
        \\
        \frac{d\bm a_I(t)}{dt}
    \end{bmatrix}
    &=-i
     \begin{bmatrix}
        \bm Z(t) & 0
        \\
        0& \bm Z^{\rm T}(t)
    \end{bmatrix} \begin{bmatrix}
        U^\dagger(t) \bm a^\dagger_I U(t)
        \\
        U^\dagger(t) \bm a^\dagger_S U(t)
    \end{bmatrix}.
\end{align}
Putting Eqs. \eqref{eq:signal trans} and \eqref{eq:idler trans} for $U^\dagger(t) \bm a_S U(t)$ and $U^\dagger(t) \bm a_I U(t)$ into Eq. \eqref{eq:heisenberg eq} above we obtain
\begin{align}
\label{eq:daSdt 2}
\frac{d\bm a_S(t)}{dt}&= -i\bm Z(t) \bm W^*_{IS}(t) \bm a_S -i\bm Z(t) \bm V^*_{II}(t)\bm a^\dagger_I,
        \\
\label{eq:daIdt 2}
\frac{d\bm a_I(t)}{dt}&=-i\bm Z^{\rm T}(t) \bm W^*_{SI}(t) \bm a_I -i\bm Z^{\rm T}(t) \bm V^*_{SS}(t)\bm a^\dagger_S.
\end{align}
Comparing Eqs. \eqref{eq:daSdt 2} and \eqref{eq:daIdt 2}  with Eqs. \eqref{eq:daSdt} and \eqref{eq:daIdt} above and equating the coefficients in front of similar operators we obtain the differential equations \eqref{eq:Vdot} and \eqref{eq:Wdot} for $d\bm V_{SS}(t)/dt$ and $d\bm V_{II}(t) / dt$, and  $d\bm W_{SI}(t)/dt$ and $d\bm W_{IS}(t) / dt$ respectively. Once we have obtained the solutions $\bm V_{SS}(t)$, $\bm V_{II}(t)$, $\bm W_{SI}(t)$, and $\bm W_{IS}(t)$ we use the results of Appendix \ref{sec:U(t)} to extract the rotation matrices $\bm \phi_S(t)$ and $\bm \phi_I(t)$ and the squeezing matrix  $\bm J(t)$ from the solutions.

To obtain the equation for the coherent state amplitudes for the pump, consider
\begin{align}
    \frac{d \bm b_P(t)}{dt} &\equiv \frac{d}{dt}\left[ U^\dagger(t) \bm b_P U(t)\right]= \bm x(t),
\end{align}
where we have introduced the column vector of pump operators $\bm b_P = [ b_{P1}, b_{P2}, \ldots]^{\rm T}$ and used Eq. \eqref{eq:pump trans} for $U^\dagger(t) \bm b_P U(t)$. In the Heisenberg picture the derivative of the pump operators using Eq. \eqref{eq:HG} for $H_{\rm G}(t)$ is given by
\begin{align}
    \frac{d \bm b_P(t)}{dt} = -i \bm \Delta(t).
\end{align}
Equating these two expressions for the pump derivative gives Eq. \eqref{eq:xdot} for $d\bm x(t)/dt$.

\section{Derivative of the non-Gaussian ket}
\label{sec:dpsidt}
In this section we derive Eq. \eqref{eq:psi tilde numerical} for the derivative of the non-Gaussian ket.

We begin by taking the  total time derivative of Eq. \eqref{eq:psi tilde numbers} for $|\tilde{\psi}(t)\rangle$ expanded in the time-dependent supermode Fock states
\begin{align}
\label{eq:psi tilde derivative}
\frac{d|\tilde{\psi}(t)\rangle}{dt}  &=\frac{\partial |\tilde{\psi}(t)\rangle}{\partial t}\nonumber
\\
&+c_{\{n_{S}\}\{n_{I}\}\{n_{P}\}}(t)\frac{d\ket{\{n_{S}\}}_t}{dt}\ket{\{n_{I}\}}_t\ket{\{n_{P}\}}_t \nonumber
\\
&+c_{\{n_{S}\}\{n_{I}\}\{n_{P}\}}(t)\ket{\{n_{S}\}}_t\frac{d\ket{\{n_{I}\}}_t}{dt}\ket{\{n_{P}\}}_t \nonumber
\\
&+c_{\{n_{S}\}\{n_{I}\}\{n_{P}\}}(t)\ket{\{n_{S}\}}_t\ket{\{n_{I}\}}_t \frac{d\ket{\{n_{P}\}}_t}{dt},
\end{align}
where we put $\{n_{S}\} = n_{S1}, \ldots, n_{SM}$, $\{n_{I}\} = n_{I1}, \ldots, n_{IM}$, and $\{n_{P}\} = n_{P1}, \ldots, n_{PL}$, and we have dropped the summation for convenience.  Here $\partial |\tilde{\psi}(t) \rangle / \partial t$ is given by Eq. \eqref{eq:partial psi tilde}.

To proceed with Eq. \eqref{eq:psi tilde derivative} we need to take the total derivative of the basis kets.  We focus on the derivative of the signal basis kets, but the derivatives of the idler and pump basis kets are derived in a similar fashion.  Using Eq. \eqref{eq:signal fock ket} for $|n_{Su}\rangle_t$ we can write
\begin{align}
\label{eq:ddt signal ket}
\frac{d\ket{\{n_S\}}_t}{dt}
&=\frac{1}{\sqrt{n_{S1}!\ldots n_{SM}!}}\nonumber
\\
&\times \bigg(\frac{d(A^\dagger_{S1})^{n_{S1}}}{dt}\ldots (A^\dagger_{SM})^{n_{SM}} \ket{0}_1\ldots\ket{0}_M \nonumber
\\
&+\ldots \nonumber
\\
&+ (A^\dagger_{S1})^{n_{S1}}\ldots \frac{d(A^\dagger_{SM})^{n_{SM}}}{dt} \ket{0}_1\ldots\ket{0}_M \bigg),
\end{align}
where we have dropped the explicit time dependence of operators for clarity and will do so throughout the remainder of this section. 
Using Eq. \eqref{eq:AS} for $A_{Su}(t)$ we can write
\begin{align}
\label{eq:ddt A}
\frac{d(A^\dagger_{Su})^{n_{Su}}}{dt} &=n_{Su} \sum_{u'}
\left[\frac{d \bm G^*_S(t)}{dt}\bm G^{\rm T}_{S}(t)\right]_{uu'}\nonumber
\\
&\times A^\dagger_{Su'} (A^{\dagger }_{Su})^{(n_{Su} -1)},
\end{align}
where $\bm G_S(t)$ is given by SVD \eqref{eq:joint svd}. Putting Eq. \eqref{eq:ddt A} into Eq. \eqref{eq:ddt signal ket}, after some straightforward algebra we obtain
\begin{align}
\label{eq:ddt signal ket 2}
\frac{d\ket{\{n_S\}}_t}{dt}
&=\sum_{u'}\left[\frac{d \bm G^*_S(t)}{dt}\bm G^{\rm T}_{S}(t)\right]_{uu'}A^\dagger_{Su'} A_{Su} \ket{\{n_S\}}_t.
\end{align}
Following similar steps we obtain
\begin{align}
\label{eq:ddt idler ket}
\frac{d\ket{\{n_I\}}_t}{dt}
&=\sum_{u'}\left[\frac{d \bm G_I(t)}{dt}\bm G^{\dagger}_{I}(t)\right]_{uu'} A^\dagger_{Iu'} A_{Iu} \ket{\{n_I\}}_t,
\\
\label{eq:ddt pump ket}
\frac{d\ket{\{n_P\}}_t}{dt}
&=\sum_{u'}\left[\frac{d \bm X^{\rm T} (t)}{dt}\bm X^*(t)\right]_{uu'} B^\dagger_{Pu'} B_{Pu} \ket{\{n_P\}}_t,
\end{align}
where $\bm G_{I}(t)$ and $\bm X(t)$ are given by SVDs \eqref{eq:joint svd} and \eqref{eq:L svd} respectively. 

Finally, we put Eq. \eqref{eq:non gaussian schrodinger equation} for $i\hbar\, d |\tilde{\psi}(t) \rangle/dt$ into the left-hand side of Eq. \eqref{eq:psi tilde derivative} and put Eqs. \eqref{eq:ddt signal ket 2} - \eqref{eq:ddt pump ket} into the right-hand side of Eq. \eqref{eq:psi tilde derivative}. Isolating the result for $i\hbar \, \partial |\tilde{\psi}(t)\rangle/\partial t$ we obtain Eq. \eqref{eq:psi tilde numerical} in the main text. 

\section{Treatment of squeezing in the Gaussian approximation using asymptotic fields}
\label{sec:lo shapes}

In this section we model a homodyne measurement of the generated squeezed state within the Gaussian approximation in the actual output channel. 
We describe the Lagrange multiplier method that we use to obtain the optimal local oscillator that minimizes the quadrature noise, and we derive an analytical formula that approximates the minimum noise. 

Consider a general quadrature operator written in terms of the asymptotic-out operators for the actual output channel
\begin{align}
\label{eq:Xac}
    X_{\rm ac} = \frac{1}{\sqrt{2}}\sum_{k}\left([\bm f_S]_k a_{{\rm ac}S k} + [\bm f_I]_{k} a_{{\rm ac}I k} \right) + {\rm H.c.},
\end{align}
where $k$ is a wavenumber in the actual waveguide, and the vectors $\bm f_S$ and $\bm f_I$ are the local oscillator profiles for the signal and idler, respectively. The local oscillator profiles for the signal and idler are normalized such that
\begin{align}
    1&=\frac{\bm f^\dagger_S \bm f_S +\bm f^\dagger_I \bm f_I}{2},
\end{align}
ensuring that the commutator is satisfied
\begin{align}
    [X_{\rm ac}, X^\dagger_{\rm ac}] = 1. 
\end{align}
The orthogonal quadrature operator $Y_{\rm ac}$ is given by
\begin{align}
    \label{eq:Yac}
  Y_{\rm ac} = -\frac{i}{\sqrt{2}}\sum_{k}\left([\bm f_S]_k a_{{\rm ac}S k} + [\bm f_I]_{k} a_{{\rm ac}I k} \right) + {\rm H.c.},
\end{align}
where
\begin{align}
\label{eq:XY comm}
    [X_{\rm ac}, Y_{\rm ac}] = i. 
\end{align}

At a very late time $t=t_1$ when the pump pulse has left the ring, and the generated signal and idler light is far from the ring propagating in the actual and phantom output channels, we consider a homodyne measurement in the actual output channel. The noise in the $X_{\rm ac}$ quadrature within the Gaussian approximation is given by
\begin{align}
\label{eq:Xnoise}
\bra{\psi(t_1)}X^2_{\rm ac}\ket{\psi(t_1)}&=\bm f^\dagger_S \bm N^{\rm ac\, ac}_{SS}(t_1)\bm f_S + \bm f^\dagger_I \bm N^{\rm ac\, ac}_{II}(t_1)\bm f_I \nonumber
\\
&+\left(\bm f^{\rm T}_S\bm M^{\rm ac\, ac}_{SI}(t_1) \bm f_I + {\rm c.c}\right)+1,
\end{align}
where $\bm M^{\rm ac\, ac}_{SI}(t_1)$ and $\bm N^{\rm ac\, ac}_{JJ}(t_1)$  are given by Eqs. \eqref{eq:M block} and \eqref{eq:N block}, respectively.   The commutation relation \eqref{eq:XY comm} between $X_{\rm ac}$ and $Y_{\rm ac}$ produces the uncertainty relation
\begin{align}
    \bra{\psi(t_1)}X^2_{\rm ac}\ket{\psi(t_1)}\bra{\psi(t_1)}Y^2_{\rm ac}\ket{\psi(t_1)} \ge 1,
\end{align}
with equality achieved for the vacuum state, which sets the vacuum noise level to $1$.
So the condition for squeezing in the $X_{\rm ac}$ quadrature below the vacuum noise is given by
\begin{align}
    \bra{\psi(t_1)}X^2_{\rm ac}\ket{\psi(t_1)} < 1.
\end{align}
The optimal local oscillator shapes $\bm f_S$ and $\bm f_I$ that minimize Eq. \eqref{eq:Xnoise} for the quadrature noise are obtained using the method of Lagrange multipliers, as described below.

Consider Eq. \eqref{eq:Xnoise} for the quadrature noise can be written without the time-dependence as
\begin{align}
\label{eq:Xnoise combined}
\langle X^2\rangle&=\bm f^{*\rm T}\bm N\bm f+\frac{1}{2}\left(\bm f^{\rm T}\bm M \bm f + {\rm c.c}\right)+\frac{1}{2}\bm f^{\rm T}\bm f^* ,
\end{align}
where we have introduced
\begin{align}
    \bm f &= \begin{bmatrix}
        \bm f_S \\ \bm f_I
    \end{bmatrix},
\\
\bm N&= \begin{bmatrix}
    \bm N^{\rm ac \, ac}_{SS} & 0 
    \\
    0 & \bm N^{\rm ac\, ac}_{II}
\end{bmatrix},
\\
\bm M&= \begin{bmatrix}
   0& \bm M^{\rm ac \, ac}_{SI}  
    \\
    \bm M^{{\rm ac \, ac} \rm T}_{SI}  & 0
\end{bmatrix},
\end{align}
and we have put $\langle X^2 \rangle = \bra{\psi(t_1)}X^2_{\rm ac}\ket{\psi(t_1)}$ for convenience. We treat the problem of minimizing the noise subject to the constraint $\frac{1}{2}\bm f^{\rm T} \bm f^*  -1=0$ using the method of Lagrange multipliers. The Lagrangian is written as
\begin{align}
    \mathcal{L}(\bm f, \bm f^*, \lambda) = \langle X^2\rangle + \lambda \left(\frac{1}{2}\bm f^{\rm T} \bm f^*  -1\right),
\end{align}
where $\lambda$ is the Lagrange multiplier. Requiring that all partial derivatives of the Lagrangian vanish leads to the following eigenvalue equation
\begin{align}
\label{eq:Lagrange eigenvalue equation}
     \begin{bmatrix}
        \Re{\bm N+\bm M}& -\Im{\bm N+\bm M} \\
        \Im{\bm N-\bm M} &  \Re{\bm N-\bm M}
    \end{bmatrix}
    \begin{bmatrix}
        \bm \sigma \\ \bm s
    \end{bmatrix}
    &=
    -\left(\lambda + \frac{1}{2}\right)
    \begin{bmatrix}
        \bm \sigma \\ \bm s
    \end{bmatrix}
\end{align}
where $\bm \sigma$ and $\bm s$ are complex vectors related to the local oscillator profiles $\bm f$ by
\begin{align}
    \bm f = \sqrt{2}\left(\bm \sigma + i \bm s \right),
\end{align}
and satisfy the normalization condition
\begin{align}
   1= \bm \sigma^{\rm T} \bm \sigma + \bm s^{\rm T} \bm s.
\end{align}
The quadrature noise is given by
\begin{align}
    \langle X^2 \rangle = -2\lambda.
\end{align}
The minimum noise corresponds to the negative $\lambda$ that is closest to zero, and the optimal local oscillator is given by the eigenvector corresponding to this $\lambda$.

The Lagrangian method is fully numerical but reliably gives the optimal local oscillators. We now derive an approximate analytical expression for the quadrature noise that closely matches the minimum noise obtained numerically.

Due to the point-coupling model we use to obtain the field components in the different channels, and the resulting partitioning of the nonlinear coefficient into a $2\times 2$ block matrix, the full Gaussian moments are given by
\begin{align}
\label{eq:M kron}
    \bm M_{SI} &= \begin{bmatrix}
        1 & \sqrt{\frac{1-\eta_I}{\eta_I}}
        \\
         \sqrt{\frac{1-\eta_S}{\eta_S}} &  \sqrt{\frac{1-\eta_I}{\eta_I}} \sqrt{\frac{1-\eta_S}{\eta_S}}
    \end{bmatrix}
    \otimes \bm M^{\rm ac\, ac}_{SI},
    \\
    \label{eq:N kron}
    \bm N_{JJ} &= \begin{bmatrix}
        1 & \sqrt{\frac{1-\eta_J}{\eta_J}}
        \\
         \sqrt{\frac{1-\eta_J}{\eta_J}} &  \frac{1-\eta_J}{\eta_J}
    \end{bmatrix}
    \otimes \bm N^{\rm ac\, ac}_{JJ},
\end{align}
where $\eta_J$ for $J=S,I$ is the escape efficiency into the actual channel and $\otimes$ is the Kronecker product. Here $\bm M_{SI}$ and $\bm N_{JJ}$ are $2n_k\times 2n_k$ matrices, where $n_k$ is the number of discrete wavenumbers used to resolve the signal or idler resonance, and $\bm M^{\rm ac\,ac}_{SI}$ and $\bm N^{\rm ac\, ac}_{JJ}$ are $n_k\times n_k$ matrices. 

Consider the SVD of the matrix partitians associated with the actual channel
\begin{align}
\label{eq:M ac svd}
    \bm M^{\rm ac\, ac}_{SI} & = \bm U_{SI} \bm \Sigma_{SI} \bm O_{SI},
    \\
\label{eq:N ac svd}
    \bm N^{\rm ac\, ac}_{JJ} & = \bm U_{JJ} \bm \Sigma_{JJ} \bm U^\dagger_{JJ},
\end{align}
where $\bm U_{JJ}$, $\bm U_{SI}$, and $\bm O_{SI}$ are $n_k \times n_k$ unitary matrices, and $\bm \Sigma_{SI}$ and $\bm \Sigma_{JJ}$ are $n_k \times n_k$ diagonal matrices. Putting Eqs. \eqref{eq:M ac svd} and \eqref{eq:N ac svd} into Eqs. \eqref{eq:M kron} and \eqref{eq:N kron} and using the mixed-product property of the Kronecker product we obtain
\begin{align}
\label{eq:M kron svd}
    \bm M_{SI}&=\left( 
   \bm Q_S
    \otimes
    \bm U_{SI}
    \right)\left( 
    \begin{bmatrix}
       \frac{1}{\sqrt{\eta_S\eta_I}} &0
        \\
        0 & 0
    \end{bmatrix}
    \otimes
    \bm \Sigma_{SI}
    \right)\left( 
    \bm Q_I
    \otimes
    \bm O_{SI}
    \right),
    \\
    \label{eq:N kron svd}
    \bm N_{JJ}&=\left( 
    \bm Q_J
    \otimes
    \bm U_{JJ}
    \right)\left( 
    \begin{bmatrix}
       \frac{1}{\eta_J} &0
        \\
        0 & 0
    \end{bmatrix}
    \otimes
    \bm \Sigma_{JJ}
    \right)\left( 
    \bm Q_J
    \otimes
    \bm U^\dagger_{JJ}
    \right),
\end{align}
where we have introduced the  $2\times 2$ unitary matrix $\bm Q_J$ as
\begin{align}
\bm Q_J = 
     \begin{bmatrix}
        -\sqrt{\eta_J} & -\sqrt{1-\eta_J}
        \\
        -\sqrt{1-\eta_J} & \sqrt{\eta_J}
    \end{bmatrix},
\end{align}
where $J=S,I$.

On the other hand the joint SVD of $\bm M_{SI}$ and $\bm N_{JJ}$ can be written as (c.f. Eq. \eqref{eq:joint svd} and Eqs. \eqref{eq:M}, \eqref{eq:NS}, and \eqref{eq:NI}) 
\begin{align}
\label{eq:M joint svd}
    \bm M_{SI} &= \bm F_S \sinh(\bm r)\cosh(\bm r)\bm F^\dagger_I,
    \\
    \label{eq:NS joint svd}
        \bm N_{SS} &= \bm F^*_S \sinh^2(\bm r)\bm F^{\rm T}_S,
          \\
    \label{eq:NI joint svd}
        \bm N_{II} &= \bm F_I \sinh^2(\bm r)\bm F^\dagger_I,
\end{align}
where $\bm r$ is a $2n_k\times 2n_k$ diagonal matrix where $ \sinh^2(\bm r) = {\rm diag}(\sinh^2(r_1), \sinh^2(r_2),\ldots)$. We compare these equations with Eqs. \eqref{eq:M kron svd}
 and \eqref{eq:N kron svd} and immediately obtain relations between the singular values
 \begin{align}
 \label{eq:M sing ac}
     \bm \Sigma_{SI}&= \sqrt{\eta_S \eta_I} \sinh(\bm r_{[1,1]})\cosh(\bm r_{[1,1]}),
     \\
  \label{eq:N sing ac}
     \bm \Sigma_{JJ}&= \eta_J \sinh^2(\bm r_{[1,1]}),    
 \end{align}
where we have defined the $n_k\times n_k$  matrix $\bm r_{[1,1]}$ as the upper left block of the full $2n_k\times 2n_k$ matrix $\bm r$.

 Having  Eqs. \eqref{eq:M sing ac} and \eqref{eq:N sing ac} for the singular values in hand, we now put Eqs. \eqref{eq:M ac svd} and \eqref{eq:N ac svd} into Eq. \eqref{eq:Xnoise} for the quadrature noise 
\begin{align}
\label{eq:Xnoise appendix}
\langle X^2 \rangle&= (\bm f^{\rm T}_{S}\bm U^*_{SS}\overline{\bm \Sigma}_{SI}\bm U^\dagger_{II}\bm f_I + {\rm c.c})\nonumber
\\
&+\bm f^{\dagger}_{S}\bm U_{SS}\bm \Sigma_{SS}\bm U^\dagger_{SS}\bm f_S  +\bm f^{\dagger}_{I}\bm U_{II}\bm \Sigma_{II}\bm U^\dagger_{II}\bm f_I + 1,
\end{align}
where we have introduced the $n_k\times n_k$ matrix
\begin{align}
    \overline{\bm \Sigma}_{SI} = \bm U^{\rm T}_{SS} \bm U_{SI} \bm \Sigma_{SI} \bm O_{SI}\bm U_{II}.
\end{align}
We make the ansatz that the minimum quadrature noise is closely achieved when the signal local oscillator $\bm f_S$ is the first column vector of $\bm U_{SS}$ and the idler local oscillator $\bm f_I$ is the first column vector of $\bm U_{II}$, i.e.,  the optimal local oscillators are approximately given by
\begin{align}
\label{eq:fS opt}
    [\bm f_S]_k &= {\rm e}^{i\theta}[\bm U_{SS}]_{k1},
    \\
\label{eq:fI opt}
    [\bm f_I]_{k} &= [\bm U_{II}]_{k1},
\end{align}
where $\theta$ is a global phase.
These vectors correspond to the largest singular values in $\bm \Sigma_{SS}$ and $\bm \Sigma_{II}$. 

Putting Eqs. \eqref{eq:fS opt} and \eqref{eq:fI opt} for the optimal local oscillators into Eq. \eqref{eq:Xnoise appendix} for the noise we obtain
\begin{align}
\label{eq:Xnoise appendix 2}
\langle X^2 \rangle&= \sqrt{\eta_S\eta_I}\big({\rm e}^{i\theta}\left[\sinh(\bm r_{[1,1]})\cosh(\bm r_{[1,1]})\right]_{uu}P_u+ {\rm c.c}\big)\nonumber
\\
&+(\eta_S + \eta_I) \sinh^2(r_1)  + 1,
\end{align}
where summation over $u$ is implied and we have introduced 
\begin{align}
   P_u = [\bm U^{\rm T}_{SI} \bm U_{SS}]_{u1}[\bm O_{SI}\bm U_{II}]_{u1}.
\end{align}
 Here $\sinh^2(r_1)$ is the number of signal and idler photon pairs corresponding to the largest singular value $r_1$. 
We choose the phase $\theta$ such that $\exp(i\theta)P_1 = -1$. Putting this into Eq. \eqref{eq:Xnoise appendix 2} we obtain
\begin{align}
\label{eq:Xnoise appendix 3}
\langle X^2 \rangle&= \frac{\exp(-2r_1)}{2}\left(\frac{\eta_S+\eta_I}{2}+\sqrt{\eta_S\eta_I}\right)\nonumber
\\
&+\frac{\exp(2r_1)}{2}\left(\frac{\eta_S+\eta_I}{2}-\sqrt{\eta_S\eta_I}\right) \nonumber
\\
&+1-\frac{\eta_S+\eta_I}{2} \nonumber
\\
&-\frac{\sqrt{\eta_S\eta_I}}{2}\sum_{u=2}^{n_k}\sinh(2r_u)\big(P^*_1P_u+ {\rm c.c}\big).
\end{align}
We find numerically that $P_u \approx 0$ for $u\ge 2$, so we can neglect the last line of Eq. \eqref{eq:Xnoise appendix 3}. So the minimum quadrature noise is approximately given by
\begin{align}
\label{eq:Xnoise appendix 4}
\langle X^2 \rangle&\approx  \frac{\exp(-2r_1)}{2}\left(\frac{\eta_S+\eta_I}{2}+\sqrt{\eta_S\eta_I}\right)\nonumber
\\
&+\frac{\exp(2r_1)}{2}\left(\frac{\eta_S+\eta_I}{2}-\sqrt{\eta_S\eta_I}\right) \nonumber
\\
&+1-\frac{\eta_S+\eta_I}{2}.
\end{align}
For $\eta_S=\eta_I =\eta$ and $\exp(-2r_1)\rightarrow 0$  the noise is simply limited by $1-\eta$.

\section{Wigner function in the actual channel for the non-Gaussian ket}
\label{sec:actual channel wigner}
In this section we show how to obtain the Wigner function of the non-Gaussian ket in the actual channel.

Consider the operators $\dd_S$ and $\dd_I$ that are linear combinations of the signal and idler operators for the asymptotic field for the actual output channel 
\begin{align}
\label{eq:ds}
    \dd_S&=\bm \beta^{\rm T}_S \bm a_{{\rm ac} S},
    \\
    \label{eq:di}
    \dd_I&=\bm \beta^{\rm T}_I \bm a_{{\rm ac} I},
\end{align}
where $\bm a_{{\rm ac} S}$ and $\bm a_{{\rm ac} I}$ are $n_k\times 1$ column vectors of operators for the $n_k$ modes in the actual channel for the signal and idler and  $\bm \beta_{S}$ and $\bm \beta_I$ are  $n_k\times 1$ column vector of the expansion coefficients that we will determine. We normalize $\bm \beta_{S}$ and $\bm \beta_I$ so that the commutators are satisfied
\begin{align}
    [\dd_S, \dd_S^\dagger]=  [\dd_I, \dd_I^\dagger] = 1.
\end{align}
It will be convenient to write Eqs. \eqref{eq:ds} and \eqref{eq:di} as
\begin{align}
\label{eq:ds 2}
    \dd_S&=\bm \beta^{\rm T}_S \begin{bmatrix}
        \bm I_{n_k} & \bm 0_{n_k}
    \end{bmatrix}\bm a_{S},
    \\
    \label{eq:di 2}
    \dd_I&=\bm \beta^{\rm T}_I\begin{bmatrix}
        \bm I_{n_k} & \bm 0_{n_k}
    \end{bmatrix}\bm a_{I},
\end{align}
where $\bm I_{n_k}$ and $\bm 0_{n_k}$ are the $n_k \times n_k$ identity matrix and zeros matrix respectively, and we have introduced the $2n_k\times 1$ column vectors $\bm a_{S} = [\bm a_{{\rm ac}S}^{\rm T}\,\,\,\bm a_{{\rm ph} S}^{\rm T}]^{\rm T}$ and $\bm a_{I} = [\bm a_{{\rm ac}I}^{\rm T}\,\,\,\bm a_{{\rm ph} I}^{\rm T}]^{\rm T}$. Here $[\bm I_{n_k} \,\,\, \bm 0_{n_k}]$
is an $n_k \times 2n_k$ matrix that projects onto the part of the operator associated with the actual channel.

We will calculate the Wigner function for the non-Gaussian ket for the actual channel in terms of $\dd_S$ and $\dd_I$. However, Eq. \eqref{eq:psi tilde numbers} for the non-Gaussian ket is written in terms of Eqs. \eqref{eq:AS} and \eqref{eq:AI} for the non-Gaussian supermode operators at the final time $\bm A_S(t_1)$ and $\bm A_I(t_1)$. It can be easily shown that Eqs. \eqref{eq:ds 2} and \eqref{eq:di 2} for $\dd_S$ and $\dd_I$ can be written in terms of $\bm A_S(t_1)$ and $\bm A_I(t_1)$ as
\begin{align}
\label{eq:ds 3}
    \dd_S&=\bm \beta^{T}_S \mathcal{G}_S\bm A_S(t_1),
    \\
    \label{eq:di 3}
    \dd_I&=\bm \beta^{T}_I \mathcal{G}_I\bm A_{I}(t_1),
\end{align}
where we have introduced the $n_k \times 2n_k$ reduced matrices $\mathcal{G}_S$ and $\mathcal{G}_I$ as
\begin{align}
\label{eq:GS script}
    \mathcal{G}_S&= \begin{bmatrix}
        \bm I_{n_k} & \bm 0_{n_k}
    \end{bmatrix} \bm G^\dagger_{S}(t_1),
    \\
    \label{eq:GI script}
    \mathcal{G}_I&= \begin{bmatrix}
        \bm I_{n_k} & \bm 0_{n_k}
    \end{bmatrix}\bm G^{\rm T}_{I}(t_1),
\end{align}
which are formed by removing the rows associated with the phantom channel from the original $2n_k \times 2n_k$  matrices $\bm G^\dagger_{S}(t_1)$ and $\bm G^{\rm T}_{I}(t_1)$. Eqs. \eqref{eq:ds 3} and \eqref{eq:di 3} for $\dd_S$ and $\dd_I$ represent a weighted sum of the supermode operators for the signal and idler, where the weights are given by the $1\times 2n_k$ row vectors $\bm \beta^{T}_S \mathcal{G}_S$  and $\bm \beta^{T}_I \mathcal{G}_I$ respectively.

We will now show that these weights contribute to a Gaussian noise term in the Wigner characteristic function. 
Consider the hybrid mode  $(\dd_S +\dd_I)/\sqrt{2}$ in the actual channel. Using Eqs. \eqref{eq:ds 3} and \eqref{eq:di 3} we can write the Wigner function for $|\tilde{\psi}(t_1)\rangle$ in terms of this hybrid mode as
\begin{widetext}
    \begin{align}
\label{eq:wigner actual channel}
    W^{\rm NG}_{\rm ac}(q,p) =  \int \frac{dxdy}{\pi^2} \langle \tilde{\psi}(t_1)|\exp(\frac{x+iy}{\sqrt{2}}[\bm \beta^\dagger_S\mathcal{G}^*_S \bm A^\dagger_S(t_1)+ \bm \beta^\dagger_I\mathcal{G}^*_I \bm A^\dagger_I(t_1)] - {\rm H.c.})|\tilde{\psi}(t_1)\rangle{\rm e}^{2i(px - qy)}.
\end{align}
\end{widetext}
Assuming that the non-Gaussian ket $|\tilde{\psi}(t_1)\rangle$ is vacuum for all modes expect the first signal and idler modes we can write Eq. \eqref{eq:wigner actual channel} as
\begin{widetext}
    \begin{align}
\label{eq:wigner actual channel 2}
    W^{\rm NG}_{\rm ac}(q,p) &=  \int \frac{dxdy}{\pi^2} \langle \tilde{\psi}(t_1)|\exp(\frac{x+iy}{\sqrt{2}}\left\{[\bm \beta^\dagger_S\mathcal{G}^*_S]_1 A^\dagger_{S1}(t_1)+ [\bm \beta^\dagger_I\mathcal{G}^*_I]_1 A^\dagger_{I1}(t_1)\right\}- {\rm H.c.})|\tilde{\psi}(t_1)\rangle \nonumber
    \\
    &\times \exp(-\frac{x^2+y^2}{4}\sum_{u=2}^{2n_k}\left\{ \left| [\bm \beta^\dagger_S\mathcal{G}^*_S]_u\right|^2 + \left|[\bm \beta^\dagger_I\mathcal{G}^*_I]_u \right|^2\right\}){\rm e}^{2i(px - qy)},
\end{align}
\end{widetext}
where $[\bm \beta^\dagger_S\mathcal{G}^*_S]_u$ and $[\bm \beta^\dagger_I\mathcal{G}^*_I]_u$ are the weights associated with the signal and idler supermode $u$. 
So the supermodes $u\ge 2$ that are not included in the ket contribute to the vacuum noise term in Eq. \eqref{eq:wigner actual channel 2}. The noise term vanishes for the ideal weights satisfying $\bm \beta^\dagger_S\mathcal{G}^*_S = \bm \beta^\dagger_I\mathcal{G}^*_I = [1 \,\,\, 0 \,\,\,\ldots\,\,\, 0 ]$, where there are $2n_k -1$ zeros and we put the first element to 1 but any non-zero complex number would also make the noise term vanish. In the presence of loss the noise term will never vanish. The matrices $\mathcal{G}_S$  and $\mathcal{G}_I$  are not invertible, so it is not possible to obtain coefficients $\bm \beta_S$ and $\bm \beta_I$ that yield the ideal weights. The best one can do is minimize the vacuum  noise term in Eq. \eqref{eq:wigner actual channel 2} by choosing the optimal coefficients $\bm \beta_S$ and $\bm \beta_I$ according to
\begin{align}
\label{eq:optimal betaS}
    \bm \beta^{\rm T}_S = \begin{bmatrix}
        1&0&\ldots &0
    \end{bmatrix}
    \mathcal{G}_S^{+},
    \\
    \label{eq:optimal betaI}
     \bm \beta^{\rm T}_I = \begin{bmatrix}
        1&0&\ldots &0
    \end{bmatrix}
    \mathcal{G}_I^{+},
\end{align}
where  $\mathcal{G}_S^{+}$ and $\mathcal{G}_I^{+}$ denote the pseudo-inverses of $\mathcal{G}_S$ and $\mathcal{G}_I$ respectively. 

Putting Eqs. \eqref{eq:optimal betaS} and \eqref{eq:optimal betaI} for the optimal coefficients into Eq. \eqref{eq:wigner actual channel 2} for the reduced Wigner function and for $\eta_S = \eta_I \equiv \eta$ we numerically find that the reduced Wigner function is well approximated by
\begin{widetext}
    \begin{align}
\label{eq:wigner actual channel 3}
    W^{\rm NG}_{\rm ac}(q,p) &\approx  \int \frac{dxdy}{\pi^2} \langle \tilde{\psi}(t_1)|\exp(\sqrt{\eta}\,\frac{x+iy}{\sqrt{2}}\left\{ A^\dagger_{S1}(t_1)+ A^\dagger_{I1}(t_1)\right\}- {\rm H.c.})|\tilde{\psi}(t_1)\rangle\exp(-(1-\eta)\,\frac{x^2+y^2}{2}){\rm e}^{2i(px - qy)}.
\end{align}
\end{widetext}
So in the approximation \eqref{eq:wigner actual channel 3} clearly the Wigner function approaches the vacuum result when $\eta \rightarrow 0$ and the noise term vanishes as $\eta \rightarrow 1$. We demonstrate in Fig. \ref{fig:wigner SI}(d) that for $\eta = 0.5$ the non-Gaussian features in the Wigner function are erased and remain visible for $\eta = 0.95$.

\section{Measuring the non-Gaussian features of the full ket.}
\label{sec:undo squeezing wigner}

In this section we show how a prospective measurement of the portion of light in the actual output channel could reveal the non-Gaussian features in the Wigner function for the full ket.

We consider the setup in Fig. \ref{fig:undo squeezing}, where the unitary $\mathcal{U}^\dagger$, represented by the red-lined box, acts only on the actual output channel. Before the box the state is $\ket{\psi} = U |\tilde{\psi}\rangle$ and after the box it becomes $\ket{\psi'} = \mathcal{U}^\dagger U |\tilde{\psi}\rangle$. Here $U$ (Eq. \eqref{eq:U}) is the Gaussian unitary already determined, and $|\tilde{\psi}\rangle$ is the residual non-Gaussian ket. By adjusting $\mathcal{U}^\dagger$ we aim to remove the Gaussian unitary so that $\ket{\psi'} \approx |\tilde{\psi}\rangle$.

The Wigner function for a mode in the actual channel $\dd$ using the full ket $\ket{\psi'}$ is given by
\begin{align}
\label{eq:wigner undo}
    W^{\rm full}_{\rm ac}(q,p) &= \int \frac{dxdy}{\pi^2} \bra{\psi'} \exp(\gamma \dd^\dagger - {\rm H.c.})\ket{\psi'}{\rm e}^{2i(px-qy)} \nonumber
    \\
    &= \int \frac{dxdy}{\pi^2} \langle \tilde{\psi}|\exp(\gamma \left[U^\dagger \mathcal{U} \dd \mathcal{U^\dagger} U\right]^\dagger - {\rm H.c.}) |\tilde{\psi}\rangle\nonumber
    \\
    &\times {\rm e}^{2i(px-qy)}
\end{align}
where we put $\gamma = x +iy$. Similar to what was done in Appendix \ref{sec:actual channel wigner}, we write $\dd$ as a superposition of the actual channel operators for the signal and idler
\begin{align}
    \dd= \bm \beta^{\rm T} \bm a_{\rm ac},
\end{align}
where $\bm \beta$ is the local oscillator shape and $\bm a_{\rm ac}$ is the column vector of signal and idler operators
\begin{align}
    \bm a_{\rm ac} = \begin{bmatrix}
        \bm a_{{\rm ac}S} \\ \bm a_{{\rm ac}I} 
    \end{bmatrix},
\end{align}
where $\bm a_{{\rm ac}S}$ and $\bm a_{{\rm ac}I}$ are introduced in Eqs. \eqref{eq:ds} and \eqref{eq:di} respectively.

To proceed with Eq. \eqref{eq:wigner undo} for the Wigner function we need to calculate the operation $U^\dagger \mathcal{U} \bm a_{\rm ac}\mathcal{U}^\dagger U$ on the actual channel operators.
We assume that unitary $\mathcal{U}^\dagger$ does the following transformation on the actual channel operators
\begin{align}
\label{eq:reduced U forward}
     \mathcal{U}^\dagger \bm a_{\rm ac}\mathcal{U}&= \bar{\bm V} \bm a_{\rm ac} + \bar{\bm W}\bm a^\dagger_{\rm ac},
     \\
    \label{eq:reduced U reverse}
     \mathcal{U}\bm a_{\rm ac}\mathcal{U}^\dagger &= \bar{\bm V}^\dagger \bm a_{\rm ac} - \bar{\bm W}^{\rm T}\bm a^\dagger_{\rm ac}
\end{align}
where  $\bar{\bm V}$ and $\bar{\bm W}$ are matrices that encode the squeezing and rotation in $\mathcal{U}$. We will determine these matrices to best remove the squeezing in $U$. We partition $\bar{\bm V}$ and $\bar{\bm W}$ into signal and idler parts as
\begin{align}
\label{eq:Vbar}
    \bar{\bm V}&=
    \begin{bmatrix}
        \Bar{\bm V}_{SS} & \bm 0
        \\
        \bm 0 &  \Bar{\bm V}_{II}
    \end{bmatrix},
    \\
    \label{eq:Wbar}
    \bar{\bm W}&=
    \begin{bmatrix}
        \bm 0&\Bar{\bm W}_{SI} 
        \\
        \Bar{\bm W}_{IS} & \bm 0
    \end{bmatrix}.
\end{align}
The commutation relations have to be preserved, yielding the following constraints
\begin{align}
    \bm I&=[\mathcal{U}\bm a_{\rm ac}\mathcal{U}^\dagger, \mathcal{U}\bm a^{\dagger \rm T}_{\rm ac}\mathcal{U}^\dagger] \rightarrow \bar{\bm V}^\dagger\bar{\bm V} - \bar{\bm W}^{\rm T}\bar{\bm W}^* = \bm I,
    \\
    \bm 0&=[\mathcal{U}\bm a_{\rm ac}\mathcal{U}^\dagger, \mathcal{U}\bm a^{\rm T}_{\rm ac}\mathcal{U}^\dagger] \rightarrow  \bar{\bm W}^{\rm T}\bar{\bm V}^* - \bar{\bm V}^{\dagger}\bar{\bm W}= \bm 0.
\end{align}

Applying the full unitary $U$ to Eq. \eqref{eq:reduced U reverse} for $\mathcal{U} \bm a_{\rm ac} \mathcal{U}^\dagger$ and using Eq. \eqref{eq:a(t)} for $U^\dagger \bm a U$ it is straightforward to obtain
\begin{align}
\label{eq:full U 2}
    U^\dagger \mathcal{U} \bm a_{\rm ac} \mathcal{U}^\dagger U &= \left( \bar{\bm V}^\dagger \bm V_{\rm ac} - \bar{\bm W}^{\rm T}  \bm W^*_{\rm ac}\right) \bm a \nonumber
    \\
    &+ \left( \bar{\bm V}^\dagger \bm W_{\rm ac} - \bar{\bm W}^{\rm T}  \bm V^*_{\rm ac}\right) \bm a^\dagger. 
\end{align}
where $\bm a = [\bm a^{\rm T}_S\,\,\, \bm a^{\rm T}_I]^{\rm T}$ is a column vector of signal and idler operators for the full system, and $\bm V_{\rm ac}$ and $\bm W_{\rm ac}$ are the reduced $\bm V$ and $\bm W$ matrices only keeping the elements that involve the actual channel. They are given by
\begin{align}
    \bm V_{\rm ac}&= \begin{bmatrix}
        [\bm V_{SS}]_{\rm ac\,\rm ac} & \bm [\bm V_{SS}]_{\rm ac\,\rm ph} & \bm 0 & \bm 0 
        \\
        \bm 0&\bm 0&[\bm V_{II}]_{\rm ac\,\rm ac} & [\bm V_{II}]_{\rm ac\,\rm ph} 
    \end{bmatrix},
    \\
    \bm W_{\rm ac}&= \begin{bmatrix}
        \bm 0&\bm 0&[\bm W_{SI}]_{\rm ac\,\rm ac} & \bm [\bm W_{SI}]_{\rm ac\,\rm ph} 
        \\
        [\bm W_{IS}]_{\rm ac\,\rm ac} &  [\bm W_{IS}]_{\rm ac\,\rm ph}&\bm 0&\bm 0
    \end{bmatrix},
\end{align}
where the frequency bin partitions $\bm V_{SS}$, $\bm V_{II}$, $\bm W_{SI}$, and $\bm W_{IS}$ encode the squeezing and rotation in the full unitary $U$ and are determined.  So we can see that when the matrices $\bar{\bm V}$ and $\bar{\bm W}$ associated with the Gaussian unitary $\mathcal{U}$ are chosen sufficiently close to $\bm V_{\rm ac}$ and $\bm W_{\rm ac}$ then $\bar{\bm V}^\dagger \bm V_{\rm ac} - \bar{\bm W}^{\rm T}  \bm W^*_{\rm ac} \approx \bm R_0$, where $\bm R_0$ is an arbitrary rotation matrix and $\bar{\bm V}^\dagger \bm W_{\rm ac} - \bar{\bm W}^{\rm T}  \bm V^*_{\rm ac} \approx \bm 0$. In this case Eq. \eqref{eq:full U 2} become approximately $U^\dagger \mathcal{U}\bm a_{\rm ac}\mathcal{U}^\dagger U \approx \bm R_0 \bm a$, where the squeezing has been removed, as required.  We choose $\bar{\bm V}$ and $\bar{\bm W}$ to be the matrices of the full system when there is no loss, as we discussed in the main text.

To calculate expectation values using the residual non-Gaussian ket, we need to relate the operators $\bm a$ on the right-hand side of Eq. \eqref{eq:full U 2} to the supermode operators. Using Eqs. \eqref{eq:AS} and \eqref{eq:AI} for the supermode operators we can write Eq. \eqref{eq:full U 2} as
\begin{align}
\label{eq:full U 3}
    U^\dagger \mathcal{U} \bm a_{\rm ac} \mathcal{U}^\dagger U &= \left( \bar{\bm V}^\dagger \bm V_{\rm ac} - \bar{\bm W}^{\rm T}  \bm W^*_{\rm ac}\right) \bm G \bm A \nonumber
    \\
    &+ \left( \bar{\bm V}^\dagger \bm W_{\rm ac} - \bar{\bm W}^{\rm T}  \bm V^*_{\rm ac}\right) \bm G^*\bm A^\dagger,
\end{align}
where $\bm A = [\bm A^{\rm T}_S\,\,\,\bm A^{\rm T}_I]^{\rm T}$, and for convenience we have introduced the matrix $\bm G$ as
\begin{align}
    \bm G = \begin{bmatrix}
        \bm G^\dagger_S & \bm 0
        \\
        \bm 0 & \bm G^{\rm T}_I
    \end{bmatrix},
\end{align}
where the determined matrices $\bm G_S$ and $\bm G_I$ come from the SVD \eqref{eq:joint svd} of the $\bm V$ and $\bm W$ matrices and are introduced in Eqs. \eqref{eq:AS} and \eqref{eq:AI} respectively.

Putting Eq. \eqref{eq:full U 3} for $U^\dagger \mathcal{U}\bm a_{\rm ac}\mathcal{U}^\dagger U$ into Eq. \eqref{eq:wigner undo} for the Wigner function we obtain
\begin{align}
  \label{eq:wigner undo 2}
    W^{\rm full}_{\rm ac}(q,p) = \int \frac{dxdy}{\pi^2} \chi(x,y){\rm e}^{2i(px-qy)}.  
\end{align}
where we have introduced the Wigner characteristic function as
\begin{align}
\label{eq:characteristic}
    \chi(x,y)&=\langle \tilde{\psi}|\exp(\gamma [\bm C \bm A + \bm D \bm A^\dagger]^\dagger - {\rm H.c.})|\tilde{\psi}\rangle,
\end{align}
and we have defined the vectors
\begin{align}
\label{eq:C coeff}
   \bm C&\equiv \bm \beta^{\rm T}\left( \bar{\bm V}^\dagger \bm V_{\rm ac} - \bar{\bm W}^{\rm T}  \bm W^*_{\rm ac}\right) \bm G,
    \\
    \label{eq:D coeff}
     \bm D&\equiv \bm \beta^{\rm T}\left( \bar{\bm V}^\dagger \bm W_{\rm ac} - \bar{\bm W}^{\rm T}  \bm V^*_{\rm ac}\right) \bm G^*.
\end{align}

Assuming the non-Gaussian ket $|\tilde{\psi}\rangle$ only contains photons in the first supermode $[\bm A]_1$ and vacuum in the remaining supermodes, we can write  Eq. \eqref{eq:characteristic} for the Wigner characteristic function as
\begin{align}
\label{eq:chi 2}
    \chi(x,y)
    &=\langle \tilde{\psi}|\exp(\gamma \{[\bm C]_1 [\bm A]_1 + [\bm D]_1 [\bm A^\dagger]_1\}^\dagger - {\rm H.c.})|\tilde{\psi} \rangle\nonumber
    \\
    &\times\prod_{l=2}^M\bra{{\rm vac}}\exp(\gamma \{[\bm C]_l [\bm A]_l + [\bm D]_l [\bm A^\dagger]_l\}^\dagger - {\rm H.c.}) \ket{\rm vac} \nonumber
    \\
    &=\langle \tilde{\psi}|\exp(\gamma \{[\bm C]_1 [\bm A]_1 + [\bm D]_1 [\bm A^\dagger]_1\}^\dagger - {\rm H.c.})|\tilde{\psi} \rangle\nonumber
    \\
    &\times \exp(-\frac{1}{2}\sum_{l=2  }^M[\gamma \bm C^\dagger - \gamma^* \bm D^{\rm T}]_l[\gamma^* \bm C - \gamma \bm D^*]_l),
\end{align}
where $M$ is the total number of supermodes. 

The exponential factor in Eq. \eqref{eq:chi 2} obscures the non-Gaussian features. The argument of the exponential factor depends on the local oscillator shape. We show below that by choosing the local oscillator shape appropriately, we can minimize the argument of the exponential  i.e., $\sum_{l=2  }^M[\gamma \bm C^\dagger - \gamma^* \bm D^{\rm T}]_l[\gamma^* \bm C - \gamma \bm D^*]_l \approx 0$. One way to achieve this is to require that Eq. \eqref{eq:C coeff}  for $\bm C$ to be a vector of zeros except for the first element, and Eq. \eqref{eq:D coeff} for $\bm D$ to be a vector of zeros. In this way the sum over the $l=2$ to $l=M$ elements would be zero exactly. So we take the target vectors for $\bm C$ and $\bm D$ are
\begin{align}
\label{eq:C target}
    \bm C_{\rm target} &= \begin{bmatrix}
        1&0& \ldots &0
    \end{bmatrix},
    \\
    \label{eq:D target}
    \bm D_{\rm target} &= \begin{bmatrix}
        0&0& \ldots &0
    \end{bmatrix}.
\end{align}
We attempt to satisfy Eq. \eqref{eq:D target} for $\bm D_{\rm target}$ by choosing $\bar{\bm W}$ and $\bar{\bm V}$ so that $\bar{\bm V}^\dagger \bm W_{\rm ac} - \bar{\bm W}^{\rm T}  \bm V^*_{\rm ac} \approx \bm 0$ as discussed above. 
We can try to satisfy Eq. \eqref{eq:C target} for $\bm C_{\rm target}$ by choosing the local oscillator shape according to
\begin{align}
\label{eq:LO shape undo}
    \bm \beta^{\rm T} = \bm C_{\rm target}\left[\left( \bar{\bm V}^\dagger \bm V_{\rm ac} - \bar{\bm W}^{\rm T}  \bm W^*_{\rm ac}\right) \bm G\right]^+,
\end{align}
where the $+$ denotes the pseudo-inverse. So after putting Eq. \eqref{eq:LO shape undo} into Eq. \eqref{eq:C coeff} we have that $\bm C \approx \bm C_{\rm target}$. Of course the resulting elements for $[\bm C]_l$ for $l\ge 2$ and  $[\bm D]_l$ for $l\ge 1$ will not be zero, but their total contribution in the argument of the exponential will be minimized.

 \end{document}